\documentclass[longauth]{aaEC}

\usepackage{graphicx}
\usepackage{natbib}
\usepackage{scalerel}
\usepackage{url}

\usepackage{txfonts}
\usepackage[pdfencoding=auto,psdextra]{hyperref}
\hypersetup{
    colorlinks=true,
    linkcolor=blue,
    filecolor=magenta,      
    urlcolor=blue,
    citecolor=blue
}
\makeatletter
\renewcommand*\aa@pageof{, page \thepage{} of \pageref*{aff103}}
\makeatother

\usepackage[utf8]{inputenc}

\usepackage[switch, modulo]{lineno}
\nolinenumbers

\usepackage{euclid}
\newcommand*{\kiloparsec}{\text{kpc}}
\newcommand*{\solarmass}{\text{\ensuremath{{M}_{\odot}}}}
\newcommand*{\solarluminosity}{\text{\ensuremath{{L}_{\odot}}}}
\newcommand{\orcid}[1]{} 
\begin{document}

\title{\Euclid: The $r_{\rm b}$-$M_\ast$ relation as a function of redshift. I. The $5\expo{9}\solarmass$ black hole in NGC\,1272\thanks{This paper is
    published on
       behalf of the Euclid Consortium}}


\author{R.~Saglia\orcid{0000-0003-0378-7032}\thanks{\email{saglia@mpe.mpg.de}}\inst{\ref{aff1},\ref{aff2}}
\and K.~Mehrgan\inst{\ref{aff2}}
\and S.~de~Nicola\orcid{0000-0001-6173-5404}\inst{\ref{aff2},\ref{aff3}}
\and J.~Thomas\inst{\ref{aff2},\ref{aff1}}
\and M.~Kluge\orcid{0000-0002-9618-2552}\inst{\ref{aff2}}
\and R.~Bender\orcid{0000-0001-7179-0626}\inst{\ref{aff2},\ref{aff1}}
\and D.~Delley\orcid{0000-0002-4958-7469}\inst{\ref{aff2}}
\and P.~Erwin\orcid{0000-0003-4588-9555}\inst{\ref{aff2}}
\and M.~Fabricius\orcid{0000-0002-7025-6058}\inst{\ref{aff2},\ref{aff1}}
\and B.~Neureiter\orcid{0000-0001-6564-9693}\inst{\ref{aff2},\ref{aff1}}
\and S.~Andreon\orcid{0000-0002-2041-8784}\inst{\ref{aff4}}
\and C.~Baccigalupi\orcid{0000-0002-8211-1630}\inst{\ref{aff5},\ref{aff6},\ref{aff7},\ref{aff8}}
\and M.~Baldi\orcid{0000-0003-4145-1943}\inst{\ref{aff9},\ref{aff10},\ref{aff11}}
\and S.~Bardelli\orcid{0000-0002-8900-0298}\inst{\ref{aff10}}
\and D.~Bonino\orcid{0000-0002-3336-9977}\inst{\ref{aff12}}
\and E.~Branchini\orcid{0000-0002-0808-6908}\inst{\ref{aff13},\ref{aff14},\ref{aff4}}
\and M.~Brescia\orcid{0000-0001-9506-5680}\inst{\ref{aff15},\ref{aff16},\ref{aff17}}
\and J.~Brinchmann\orcid{0000-0003-4359-8797}\inst{\ref{aff18},\ref{aff19}}
\and A.~Caillat\inst{\ref{aff20}}
\and S.~Camera\orcid{0000-0003-3399-3574}\inst{\ref{aff21},\ref{aff22},\ref{aff12}}
\and V.~Capobianco\orcid{0000-0002-3309-7692}\inst{\ref{aff12}}
\and C.~Carbone\orcid{0000-0003-0125-3563}\inst{\ref{aff23}}
\and J.~Carretero\orcid{0000-0002-3130-0204}\inst{\ref{aff24},\ref{aff25}}
\and S.~Casas\orcid{0000-0002-4751-5138}\inst{\ref{aff26},\ref{aff27}}
\and M.~Castellano\orcid{0000-0001-9875-8263}\inst{\ref{aff28}}
\and G.~Castignani\orcid{0000-0001-6831-0687}\inst{\ref{aff10}}
\and S.~Cavuoti\orcid{0000-0002-3787-4196}\inst{\ref{aff16},\ref{aff17}}
\and A.~Cimatti\inst{\ref{aff29}}
\and C.~Colodro-Conde\inst{\ref{aff30}}
\and G.~Congedo\orcid{0000-0003-2508-0046}\inst{\ref{aff31}}
\and C.~J.~Conselice\orcid{0000-0003-1949-7638}\inst{\ref{aff32}}
\and L.~Conversi\orcid{0000-0002-6710-8476}\inst{\ref{aff33},\ref{aff34}}
\and Y.~Copin\orcid{0000-0002-5317-7518}\inst{\ref{aff35}}
\and F.~Courbin\orcid{0000-0003-0758-6510}\inst{\ref{aff36},\ref{aff37},\ref{aff38}}
\and H.~M.~Courtois\orcid{0000-0003-0509-1776}\inst{\ref{aff39}}
\and H.~Degaudenzi\orcid{0000-0002-5887-6799}\inst{\ref{aff40}}
\and G.~De~Lucia\orcid{0000-0002-6220-9104}\inst{\ref{aff6}}
\and J.~Dinis\orcid{0000-0001-5075-1601}\inst{\ref{aff41},\ref{aff42}}
\and X.~Dupac\inst{\ref{aff34}}
\and S.~Dusini\orcid{0000-0002-1128-0664}\inst{\ref{aff43}}
\and M.~Farina\orcid{0000-0002-3089-7846}\inst{\ref{aff44}}
\and S.~Farrens\orcid{0000-0002-9594-9387}\inst{\ref{aff45}}
\and F.~Faustini\orcid{0000-0001-6274-5145}\inst{\ref{aff46},\ref{aff28}}
\and S.~Ferriol\inst{\ref{aff35}}
\and N.~Fourmanoit\orcid{0009-0005-6816-6925}\inst{\ref{aff47}}
\and M.~Frailis\orcid{0000-0002-7400-2135}\inst{\ref{aff6}}
\and E.~Franceschi\orcid{0000-0002-0585-6591}\inst{\ref{aff10}}
\and M.~Fumana\orcid{0000-0001-6787-5950}\inst{\ref{aff23}}
\and S.~Galeotta\orcid{0000-0002-3748-5115}\inst{\ref{aff6}}
\and K.~George\orcid{0000-0002-1734-8455}\inst{\ref{aff1}}
\and B.~Gillis\orcid{0000-0002-4478-1270}\inst{\ref{aff31}}
\and C.~Giocoli\orcid{0000-0002-9590-7961}\inst{\ref{aff10},\ref{aff48}}
\and A.~Grazian\orcid{0000-0002-5688-0663}\inst{\ref{aff49}}
\and F.~Grupp\inst{\ref{aff2},\ref{aff1}}
\and L.~Guzzo\orcid{0000-0001-8264-5192}\inst{\ref{aff50},\ref{aff4}}
\and S.~V.~H.~Haugan\orcid{0000-0001-9648-7260}\inst{\ref{aff51}}
\and J.~Hoar\inst{\ref{aff34}}
\and W.~Holmes\inst{\ref{aff52}}
\and F.~Hormuth\inst{\ref{aff53}}
\and A.~Hornstrup\orcid{0000-0002-3363-0936}\inst{\ref{aff54},\ref{aff55}}
\and K.~Jahnke\orcid{0000-0003-3804-2137}\inst{\ref{aff56}}
\and M.~Jhabvala\inst{\ref{aff57}}
\and E.~Keih\"anen\orcid{0000-0003-1804-7715}\inst{\ref{aff58}}
\and S.~Kermiche\orcid{0000-0002-0302-5735}\inst{\ref{aff47}}
\and A.~Kiessling\orcid{0000-0002-2590-1273}\inst{\ref{aff52}}
\and M.~Kilbinger\orcid{0000-0001-9513-7138}\inst{\ref{aff45}}
\and B.~Kubik\orcid{0009-0006-5823-4880}\inst{\ref{aff35}}
\and M.~K\"ummel\orcid{0000-0003-2791-2117}\inst{\ref{aff1}}
\and M.~Kunz\orcid{0000-0002-3052-7394}\inst{\ref{aff59}}
\and H.~Kurki-Suonio\orcid{0000-0002-4618-3063}\inst{\ref{aff60},\ref{aff61}}
\and D.~Le~Mignant\orcid{0000-0002-5339-5515}\inst{\ref{aff20}}
\and S.~Ligori\orcid{0000-0003-4172-4606}\inst{\ref{aff12}}
\and P.~B.~Lilje\orcid{0000-0003-4324-7794}\inst{\ref{aff51}}
\and V.~Lindholm\orcid{0000-0003-2317-5471}\inst{\ref{aff60},\ref{aff61}}
\and I.~Lloro\inst{\ref{aff62}}
\and G.~Mainetti\orcid{0000-0003-2384-2377}\inst{\ref{aff63}}
\and E.~Maiorano\orcid{0000-0003-2593-4355}\inst{\ref{aff10}}
\and O.~Mansutti\orcid{0000-0001-5758-4658}\inst{\ref{aff6}}
\and O.~Marggraf\orcid{0000-0001-7242-3852}\inst{\ref{aff64}}
\and K.~Markovic\orcid{0000-0001-6764-073X}\inst{\ref{aff52}}
\and M.~Martinelli\orcid{0000-0002-6943-7732}\inst{\ref{aff28},\ref{aff65}}
\and N.~Martinet\orcid{0000-0003-2786-7790}\inst{\ref{aff20}}
\and F.~Marulli\orcid{0000-0002-8850-0303}\inst{\ref{aff66},\ref{aff10},\ref{aff11}}
\and R.~Massey\orcid{0000-0002-6085-3780}\inst{\ref{aff67}}
\and E.~Medinaceli\orcid{0000-0002-4040-7783}\inst{\ref{aff10}}
\and M.~Melchior\inst{\ref{aff68}}
\and Y.~Mellier\inst{\ref{aff69},\ref{aff70}}
\and M.~Meneghetti\orcid{0000-0003-1225-7084}\inst{\ref{aff10},\ref{aff11}}
\and E.~Merlin\orcid{0000-0001-6870-8900}\inst{\ref{aff28}}
\and G.~Meylan\inst{\ref{aff36}}
\and M.~Moresco\orcid{0000-0002-7616-7136}\inst{\ref{aff66},\ref{aff10}}
\and L.~Moscardini\orcid{0000-0002-3473-6716}\inst{\ref{aff66},\ref{aff10},\ref{aff11}}
\and E.~Munari\orcid{0000-0002-1751-5946}\inst{\ref{aff6},\ref{aff5}}
\and R.~Nakajima\inst{\ref{aff64}}
\and C.~Neissner\orcid{0000-0001-8524-4968}\inst{\ref{aff71},\ref{aff25}}
\and R.~C.~Nichol\orcid{0000-0003-0939-6518}\inst{\ref{aff72}}
\and S.-M.~Niemi\inst{\ref{aff73}}
\and J.~W.~Nightingale\orcid{0000-0002-8987-7401}\inst{\ref{aff74}}
\and C.~Padilla\orcid{0000-0001-7951-0166}\inst{\ref{aff71}}
\and S.~Paltani\orcid{0000-0002-8108-9179}\inst{\ref{aff40}}
\and F.~Pasian\orcid{0000-0002-4869-3227}\inst{\ref{aff6}}
\and K.~Pedersen\inst{\ref{aff75}}
\and W.~J.~Percival\orcid{0000-0002-0644-5727}\inst{\ref{aff76},\ref{aff77},\ref{aff78}}
\and V.~Pettorino\inst{\ref{aff73}}
\and S.~Pires\orcid{0000-0002-0249-2104}\inst{\ref{aff45}}
\and G.~Polenta\orcid{0000-0003-4067-9196}\inst{\ref{aff46}}
\and M.~Poncet\inst{\ref{aff79}}
\and L.~A.~Popa\inst{\ref{aff80}}
\and L.~Pozzetti\orcid{0000-0001-7085-0412}\inst{\ref{aff10}}
\and F.~Raison\orcid{0000-0002-7819-6918}\inst{\ref{aff2}}
\and R.~Rebolo\inst{\ref{aff30},\ref{aff81},\ref{aff82}}
\and A.~Renzi\orcid{0000-0001-9856-1970}\inst{\ref{aff83},\ref{aff43}}
\and J.~Rhodes\orcid{0000-0002-4485-8549}\inst{\ref{aff52}}
\and G.~Riccio\inst{\ref{aff16}}
\and E.~Romelli\orcid{0000-0003-3069-9222}\inst{\ref{aff6}}
\and M.~Roncarelli\orcid{0000-0001-9587-7822}\inst{\ref{aff10}}
\and E.~Rossetti\orcid{0000-0003-0238-4047}\inst{\ref{aff9}}
\and Z.~Sakr\orcid{0000-0002-4823-3757}\inst{\ref{aff84},\ref{aff85},\ref{aff86}}
\and A.~G.~S\'anchez\orcid{0000-0003-1198-831X}\inst{\ref{aff2}}
\and D.~Sapone\orcid{0000-0001-7089-4503}\inst{\ref{aff87}}
\and B.~Sartoris\orcid{0000-0003-1337-5269}\inst{\ref{aff1},\ref{aff6}}
\and M.~Schirmer\orcid{0000-0003-2568-9994}\inst{\ref{aff56}}
\and P.~Schneider\orcid{0000-0001-8561-2679}\inst{\ref{aff64}}
\and T.~Schrabback\orcid{0000-0002-6987-7834}\inst{\ref{aff88}}
\and A.~Secroun\orcid{0000-0003-0505-3710}\inst{\ref{aff47}}
\and M.~Seiffert\orcid{0000-0002-7536-9393}\inst{\ref{aff52}}
\and S.~Serrano\orcid{0000-0002-0211-2861}\inst{\ref{aff89},\ref{aff90},\ref{aff91}}
\and C.~Sirignano\orcid{0000-0002-0995-7146}\inst{\ref{aff83},\ref{aff43}}
\and G.~Sirri\orcid{0000-0003-2626-2853}\inst{\ref{aff11}}
\and J.~Skottfelt\orcid{0000-0003-1310-8283}\inst{\ref{aff92}}
\and L.~Stanco\orcid{0000-0002-9706-5104}\inst{\ref{aff43}}
\and J.~Steinwagner\orcid{0000-0001-7443-1047}\inst{\ref{aff2}}
\and P.~Tallada-Cresp\'{i}\orcid{0000-0002-1336-8328}\inst{\ref{aff24},\ref{aff25}}
\and D.~Tavagnacco\orcid{0000-0001-7475-9894}\inst{\ref{aff6}}
\and A.~N.~Taylor\inst{\ref{aff31}}
\and I.~Tereno\inst{\ref{aff41},\ref{aff93}}
\and R.~Toledo-Moreo\orcid{0000-0002-2997-4859}\inst{\ref{aff94}}
\and F.~Torradeflot\orcid{0000-0003-1160-1517}\inst{\ref{aff25},\ref{aff24}}
\and I.~Tutusaus\orcid{0000-0002-3199-0399}\inst{\ref{aff85}}
\and L.~Valenziano\orcid{0000-0002-1170-0104}\inst{\ref{aff10},\ref{aff95}}
\and T.~Vassallo\orcid{0000-0001-6512-6358}\inst{\ref{aff1},\ref{aff6}}
\and G.~Verdoes~Kleijn\orcid{0000-0001-5803-2580}\inst{\ref{aff96}}
\and Y.~Wang\orcid{0000-0002-4749-2984}\inst{\ref{aff97}}
\and J.~Weller\orcid{0000-0002-8282-2010}\inst{\ref{aff1},\ref{aff2}}
\and G.~Zamorani\orcid{0000-0002-2318-301X}\inst{\ref{aff10}}
\and E.~Zucca\orcid{0000-0002-5845-8132}\inst{\ref{aff10}}
\and C.~Burigana\orcid{0000-0002-3005-5796}\inst{\ref{aff98},\ref{aff95}}
\and V.~Scottez\inst{\ref{aff69},\ref{aff99}}
\and L.~Ferrarese\orcid{0000-0002-8224-1128}\inst{\ref{aff100}}
\and E.~Lusso\orcid{0000-0003-0083-1157}\inst{\ref{aff101},\ref{aff102}}
\and D.~Scott\orcid{0000-0002-6878-9840}\inst{\ref{aff103}}}
										   
\institute{Universit\"ats-Sternwarte M\"unchen, Fakult\"at f\"ur Physik, Ludwig-Maximilians-Universit\"at M\"unchen, Scheinerstrasse 1, 81679 M\"unchen, Germany\label{aff1}
\and
Max Planck Institute for Extraterrestrial Physics, Giessenbachstr. 1, 85748 Garching, Germany\label{aff2}
\and
Ludwig-Maximilians-University, Schellingstrasse 4, 80799 Munich, Germany\label{aff3}
\and
INAF-Osservatorio Astronomico di Brera, Via Brera 28, 20122 Milano, Italy\label{aff4}
\and
IFPU, Institute for Fundamental Physics of the Universe, via Beirut 2, 34151 Trieste, Italy\label{aff5}
\and
INAF-Osservatorio Astronomico di Trieste, Via G. B. Tiepolo 11, 34143 Trieste, Italy\label{aff6}
\and
INFN, Sezione di Trieste, Via Valerio 2, 34127 Trieste TS, Italy\label{aff7}
\and
SISSA, International School for Advanced Studies, Via Bonomea 265, 34136 Trieste TS, Italy\label{aff8}
\and
Dipartimento di Fisica e Astronomia, Universit\`a di Bologna, Via Gobetti 93/2, 40129 Bologna, Italy\label{aff9}
\and
INAF-Osservatorio di Astrofisica e Scienza dello Spazio di Bologna, Via Piero Gobetti 93/3, 40129 Bologna, Italy\label{aff10}
\and
INFN-Sezione di Bologna, Viale Berti Pichat 6/2, 40127 Bologna, Italy\label{aff11}
\and
INAF-Osservatorio Astrofisico di Torino, Via Osservatorio 20, 10025 Pino Torinese (TO), Italy\label{aff12}
\and
Dipartimento di Fisica, Universit\`a di Genova, Via Dodecaneso 33, 16146, Genova, Italy\label{aff13}
\and
INFN-Sezione di Genova, Via Dodecaneso 33, 16146, Genova, Italy\label{aff14}
\and
Department of Physics "E. Pancini", University Federico II, Via Cinthia 6, 80126, Napoli, Italy\label{aff15}
\and
INAF-Osservatorio Astronomico di Capodimonte, Via Moiariello 16, 80131 Napoli, Italy\label{aff16}
\and
INFN section of Naples, Via Cinthia 6, 80126, Napoli, Italy\label{aff17}
\and
Instituto de Astrof\'isica e Ci\^encias do Espa\c{c}o, Universidade do Porto, CAUP, Rua das Estrelas, PT4150-762 Porto, Portugal\label{aff18}
\and
Faculdade de Ci\^encias da Universidade do Porto, Rua do Campo de Alegre, 4150-007 Porto, Portugal\label{aff19}
\and
Aix-Marseille Universit\'e, CNRS, CNES, LAM, Marseille, France\label{aff20}
\and
Dipartimento di Fisica, Universit\`a degli Studi di Torino, Via P. Giuria 1, 10125 Torino, Italy\label{aff21}
\and
INFN-Sezione di Torino, Via P. Giuria 1, 10125 Torino, Italy\label{aff22}
\and
INAF-IASF Milano, Via Alfonso Corti 12, 20133 Milano, Italy\label{aff23}
\and
Centro de Investigaciones Energ\'eticas, Medioambientales y Tecnol\'ogicas (CIEMAT), Avenida Complutense 40, 28040 Madrid, Spain\label{aff24}
\and
Port d'Informaci\'{o} Cient\'{i}fica, Campus UAB, C. Albareda s/n, 08193 Bellaterra (Barcelona), Spain\label{aff25}
\and
Institute for Theoretical Particle Physics and Cosmology (TTK), RWTH Aachen University, 52056 Aachen, Germany\label{aff26}
\and
Institute of Cosmology and Gravitation, University of Portsmouth, Portsmouth PO1 3FX, UK\label{aff27}
\and
INAF-Osservatorio Astronomico di Roma, Via Frascati 33, 00078 Monteporzio Catone, Italy\label{aff28}
\and
Dipartimento di Fisica e Astronomia "Augusto Righi" - Alma Mater Studiorum Universit\`a di Bologna, Viale Berti Pichat 6/2, 40127 Bologna, Italy\label{aff29}
\and
Instituto de Astrof\'isica de Canarias, Calle V\'ia L\'actea s/n, 38204, San Crist\'obal de La Laguna, Tenerife, Spain\label{aff30}
\and
Institute for Astronomy, University of Edinburgh, Royal Observatory, Blackford Hill, Edinburgh EH9 3HJ, UK\label{aff31}
\and
Jodrell Bank Centre for Astrophysics, Department of Physics and Astronomy, University of Manchester, Oxford Road, Manchester M13 9PL, UK\label{aff32}
\and
European Space Agency/ESRIN, Largo Galileo Galilei 1, 00044 Frascati, Roma, Italy\label{aff33}
\and
ESAC/ESA, Camino Bajo del Castillo, s/n., Urb. Villafranca del Castillo, 28692 Villanueva de la Ca\~nada, Madrid, Spain\label{aff34}
\and
Universit\'e Claude Bernard Lyon 1, CNRS/IN2P3, IP2I Lyon, UMR 5822, Villeurbanne, F-69100, France\label{aff35}
\and
Institute of Physics, Laboratory of Astrophysics, Ecole Polytechnique F\'ed\'erale de Lausanne (EPFL), Observatoire de Sauverny, 1290 Versoix, Switzerland\label{aff36}
\and
Institut de Ci\`{e}ncies del Cosmos (ICCUB), Universitat de Barcelona (IEEC-UB), Mart\'{i} i Franqu\`{e}s 1, 08028 Barcelona, Spain\label{aff37}
\and
Instituci\'o Catalana de Recerca i Estudis Avan\c{c}ats (ICREA), Passeig de Llu\'{\i}s Companys 23, 08010 Barcelona, Spain\label{aff38}
\and
UCB Lyon 1, CNRS/IN2P3, IUF, IP2I Lyon, 4 rue Enrico Fermi, 69622 Villeurbanne, France\label{aff39}
\and
Department of Astronomy, University of Geneva, ch. d'Ecogia 16, 1290 Versoix, Switzerland\label{aff40}
\and
Departamento de F\'isica, Faculdade de Ci\^encias, Universidade de Lisboa, Edif\'icio C8, Campo Grande, PT1749-016 Lisboa, Portugal\label{aff41}
\and
Instituto de Astrof\'isica e Ci\^encias do Espa\c{c}o, Faculdade de Ci\^encias, Universidade de Lisboa, Campo Grande, 1749-016 Lisboa, Portugal\label{aff42}
\and
INFN-Padova, Via Marzolo 8, 35131 Padova, Italy\label{aff43}
\and
INAF-Istituto di Astrofisica e Planetologia Spaziali, via del Fosso del Cavaliere, 100, 00100 Roma, Italy\label{aff44}
\and
Universit\'e Paris-Saclay, Universit\'e Paris Cit\'e, CEA, CNRS, AIM, 91191, Gif-sur-Yvette, France\label{aff45}
\and
Space Science Data Center, Italian Space Agency, via del Politecnico snc, 00133 Roma, Italy\label{aff46}
\and
Aix-Marseille Universit\'e, CNRS/IN2P3, CPPM, Marseille, France\label{aff47}
\and
Istituto Nazionale di Fisica Nucleare, Sezione di Bologna, Via Irnerio 46, 40126 Bologna, Italy\label{aff48}
\and
INAF-Osservatorio Astronomico di Padova, Via dell'Osservatorio 5, 35122 Padova, Italy\label{aff49}
\and
Dipartimento di Fisica "Aldo Pontremoli", Universit\`a degli Studi di Milano, Via Celoria 16, 20133 Milano, Italy\label{aff50}
\and
Institute of Theoretical Astrophysics, University of Oslo, P.O. Box 1029 Blindern, 0315 Oslo, Norway\label{aff51}
\and
Jet Propulsion Laboratory, California Institute of Technology, 4800 Oak Grove Drive, Pasadena, CA, 91109, USA\label{aff52}
\and
Felix Hormuth Engineering, Goethestr. 17, 69181 Leimen, Germany\label{aff53}
\and
Technical University of Denmark, Elektrovej 327, 2800 Kgs. Lyngby, Denmark\label{aff54}
\and
Cosmic Dawn Center (DAWN), Denmark\label{aff55}
\and
Max-Planck-Institut f\"ur Astronomie, K\"onigstuhl 17, 69117 Heidelberg, Germany\label{aff56}
\and
NASA Goddard Space Flight Center, Greenbelt, MD 20771, USA\label{aff57}
\and
Department of Physics and Helsinki Institute of Physics, Gustaf H\"allstr\"omin katu 2, 00014 University of Helsinki, Finland\label{aff58}
\and
Universit\'e de Gen\`eve, D\'epartement de Physique Th\'eorique and Centre for Astroparticle Physics, 24 quai Ernest-Ansermet, CH-1211 Gen\`eve 4, Switzerland\label{aff59}
\and
Department of Physics, P.O. Box 64, 00014 University of Helsinki, Finland\label{aff60}
\and
Helsinki Institute of Physics, Gustaf H{\"a}llstr{\"o}min katu 2, University of Helsinki, Helsinki, Finland\label{aff61}
\and
NOVA optical infrared instrumentation group at ASTRON, Oude Hoogeveensedijk 4, 7991PD, Dwingeloo, The Netherlands\label{aff62}
\and
Centre de Calcul de l'IN2P3/CNRS, 21 avenue Pierre de Coubertin 69627 Villeurbanne Cedex, France\label{aff63}
\and
Universit\"at Bonn, Argelander-Institut f\"ur Astronomie, Auf dem H\"ugel 71, 53121 Bonn, Germany\label{aff64}
\and
INFN-Sezione di Roma, Piazzale Aldo Moro, 2 - c/o Dipartimento di Fisica, Edificio G. Marconi, 00185 Roma, Italy\label{aff65}
\and
Dipartimento di Fisica e Astronomia "Augusto Righi" - Alma Mater Studiorum Universit\`a di Bologna, via Piero Gobetti 93/2, 40129 Bologna, Italy\label{aff66}
\and
Department of Physics, Institute for Computational Cosmology, Durham University, South Road, DH1 3LE, UK\label{aff67}
\and
University of Applied Sciences and Arts of Northwestern Switzerland, School of Engineering, 5210 Windisch, Switzerland\label{aff68}
\and
Institut d'Astrophysique de Paris, 98bis Boulevard Arago, 75014, Paris, France\label{aff69}
\and
Institut d'Astrophysique de Paris, UMR 7095, CNRS, and Sorbonne Universit\'e, 98 bis boulevard Arago, 75014 Paris, France\label{aff70}
\and
Institut de F\'{i}sica d'Altes Energies (IFAE), The Barcelona Institute of Science and Technology, Campus UAB, 08193 Bellaterra (Barcelona), Spain\label{aff71}
\and
School of Mathematics and Physics, University of Surrey, Guildford, Surrey, GU2 7XH, UK\label{aff72}
\and
European Space Agency/ESTEC, Keplerlaan 1, 2201 AZ Noordwijk, The Netherlands\label{aff73}
\and
School of Mathematics, Statistics and Physics, Newcastle University, Herschel Building, Newcastle-upon-Tyne, NE1 7RU, UK\label{aff74}
\and
DARK, Niels Bohr Institute, University of Copenhagen, Jagtvej 155, 2200 Copenhagen, Denmark\label{aff75}
\and
Waterloo Centre for Astrophysics, University of Waterloo, Waterloo, Ontario N2L 3G1, Canada\label{aff76}
\and
Department of Physics and Astronomy, University of Waterloo, Waterloo, Ontario N2L 3G1, Canada\label{aff77}
\and
Perimeter Institute for Theoretical Physics, Waterloo, Ontario N2L 2Y5, Canada\label{aff78}
\and
Centre National d'Etudes Spatiales -- Centre spatial de Toulouse, 18 avenue Edouard Belin, 31401 Toulouse Cedex 9, France\label{aff79}
\and
Institute of Space Science, Str. Atomistilor, nr. 409 M\u{a}gurele, Ilfov, 077125, Romania\label{aff80}
\and
Departamento de Astrof\'isica, Universidad de La Laguna, 38206, La Laguna, Tenerife, Spain\label{aff81}
\and
Consejo Superior de Investigaciones Cientificas, Calle Serrano 117, 28006 Madrid, Spain\label{aff82}
\and
Dipartimento di Fisica e Astronomia "G. Galilei", Universit\`a di Padova, Via Marzolo 8, 35131 Padova, Italy\label{aff83}
\and
Institut f\"ur Theoretische Physik, University of Heidelberg, Philosophenweg 16, 69120 Heidelberg, Germany\label{aff84}
\and
Institut de Recherche en Astrophysique et Plan\'etologie (IRAP), Universit\'e de Toulouse, CNRS, UPS, CNES, 14 Av. Edouard Belin, 31400 Toulouse, France\label{aff85}
\and
Universit\'e St Joseph; Faculty of Sciences, Beirut, Lebanon\label{aff86}
\and
Departamento de F\'isica, FCFM, Universidad de Chile, Blanco Encalada 2008, Santiago, Chile\label{aff87}
\and
Universit\"at Innsbruck, Institut f\"ur Astro- und Teilchenphysik, Technikerstr. 25/8, 6020 Innsbruck, Austria\label{aff88}
\and
Institut d'Estudis Espacials de Catalunya (IEEC),  Edifici RDIT, Campus UPC, 08860 Castelldefels, Barcelona, Spain\label{aff89}
\and
Satlantis, University Science Park, Sede Bld 48940, Leioa-Bilbao, Spain\label{aff90}
\and
Institute of Space Sciences (ICE, CSIC), Campus UAB, Carrer de Can Magrans, s/n, 08193 Barcelona, Spain\label{aff91}
\and
Centre for Electronic Imaging, Open University, Walton Hall, Milton Keynes, MK7~6AA, UK\label{aff92}
\and
Instituto de Astrof\'isica e Ci\^encias do Espa\c{c}o, Faculdade de Ci\^encias, Universidade de Lisboa, Tapada da Ajuda, 1349-018 Lisboa, Portugal\label{aff93}
\and
Universidad Polit\'ecnica de Cartagena, Departamento de Electr\'onica y Tecnolog\'ia de Computadoras,  Plaza del Hospital 1, 30202 Cartagena, Spain\label{aff94}
\and
INFN-Bologna, Via Irnerio 46, 40126 Bologna, Italy\label{aff95}
\and
Kapteyn Astronomical Institute, University of Groningen, PO Box 800, 9700 AV Groningen, The Netherlands\label{aff96}
\and
Infrared Processing and Analysis Center, California Institute of Technology, Pasadena, CA 91125, USA\label{aff97}
\and
INAF, Istituto di Radioastronomia, Via Piero Gobetti 101, 40129 Bologna, Italy\label{aff98}
\and
Junia, EPA department, 41 Bd Vauban, 59800 Lille, France\label{aff99}
\and
NRC Herzberg, 5071 West Saanich Rd, Victoria, BC V9E 2E7, Canada\label{aff100}
\and
INAF-Osservatorio Astrofisico di Arcetri, Largo E. Fermi 5, 50125, Firenze, Italy\label{aff101}
\and
Dipartimento di Fisica e Astronomia, Universit\`{a} di Firenze, via G. Sansone 1, 50019 Sesto Fiorentino, Firenze, Italy\label{aff102}
\and
Department of Physics and Astronomy, University of British Columbia, Vancouver, BC V6T 1Z1, Canada\label{aff103}}    

\date{}

 
  \abstract{
    Core ellipticals, massive
    early-type galaxies with almost constant inner surface brightness
    profiles, are the results of dry mergers. During these events a
    binary black hole is formed, destroying the original cuspy central
    regions of the merging objects and scattering stars that are not on
    tangential orbits. The size of the emerging core correlates with
    the mass of the finally merged black hole. Therefore, the
    determination of the size of the core of massive early type
    galaxies provides key insights not only on the mass of the black
    hole, but also on the origin and evolution of these objects.
    In this work we report the first
    \Euclid-based dynamical mass determination of a supermassive black
    hole.  We perform it by studying the centre of NGC 1272, the
    second most luminous elliptical galaxy in the Perseus cluster,
    combining the \Euclid VIS photometry coming from the Early Release
    Observations of the Perseus cluster with VIRUS spectroscopic
    observations at the Hobby-Eberly Telescope.
    The core of NGC 1272 is detected
    on the \Euclid VIS image. Its size is
    $1\arcsecf29\pm 0\arcsecf07$ or 0.45 \kiloparsec, determined by
    fitting PSF-convolved core-S\'ersic and Nuker-law functions. We
    deproject the surface brightness profile of the galaxy, finding
    that the galaxy is axisymmetric and nearly spherical. The
    two-dimensional stellar kinematics of the galaxy is measured from
    the VIRUS spectra by deriving optimally regularized non-parametric
    line-of-sight velocity distributions. Dynamical models of the
    galaxy are constructed using our axisymmetric and triaxial
    Schwarzschild codes.
    We measure a black hole mass of $(5\pm3)\expo{9}\solarmass$,
    in line with the expectation from the
    $M_{\rm BH}$-$r_{\rm b}$ correlation, but eight times larger than
    predicted by the $M_{\rm BH}$-$\sigma$ correlation (at $1.8\sigma$ significance).
    The core size, rather than the velocity dispersion, allows one to
    select galaxies harboring the most massive black holes. The
    spatial resolution, wide area coverage, and depth of the \Euclid
    (Wide and Deep) surveys allow us to find cores of passive galaxies
    larger than 2 \kiloparsec\ up to redshift 1. }
  \keywords{Galaxies:
    kinematics and dynamics; elliptical and lenticular, cD;
    individual: NGC 1272; nuclei; photometry }
  \titlerunning{The
    $5\expo{9}\solarmass$ black hole of NGC 1272 }
  \authorrunning{Saglia et al. }

   \maketitle
%

\section{Introduction}
\label{Sec_Intro}
Massive early-type galaxies (ETGs) are commonly found at the centre of
galaxy clusters and are the result of mostly dissipationless
mergers. During these events nuclear supermassive black hole (SMBH)
binaries are formed. Gravitational slingshots eject stars on radial
orbits from the center of the remnant galaxy, destroying the power-law
surface brightness distributions found in lower luminosity ellipticals
\citep{Faber1997}. Gravitational wave recoil \citep{Khonji2024} can
enhance the scouring mechanism. Through this core-scouring mechanism
the surface brightness profile $I(r)$ of most massive ETGs becomes
almost constant within a break (or core) radius $r_{\rm b}$, and for
$r\le r_{\rm b}$ one finds $I(r)\propto r^{-\gamma}$, with
$\gamma<0.3$ \citep{Faber1997}.  The break radius $r_{\rm b}$ is
tightly correlated with the mass of the central black hole
\citep{Rusli2013,Thomas2016}, and anti-correlated with the central
surface brightness \citep{Mehrgan2019}. A broader correlation between
$r_{\rm b}$ and luminosity or stellar mass of the galaxies is also
established \citep{Laine2003,Rusli2013}. Moreover, within $r_{\rm b}$
the distribution of orbits becomes tangentially anisotropic
\citep{Thomas2014}, as only stars avoiding the center can survive the
scouring
\citep{Milosavlijevic2001,Thomas2014,Rantala2018,Rantala2019}.
Although alternative explanations for the formation of cores have been
proposed, such as the ``tidal deposition'' discussed by
\citet{Nasim2021} and the feedback by active galactic nuclei, see
\citet{Teyssier2011}, \citet{Martizzi2012}, and \citet{Choi2018},
they fail to explain this tangential anisotropy signature.

Black Holes (BHs) with dynamically measured masses larger than
$10^{10} \solarmass$, i.e., hypermassive BHs or HMBHs, are
still rare. They cannot be found using the $M_{\rm BH}$-$\sigma$
relation \citep{Saglia2016}: a dissipationless merger of equal mass
galaxies doubles the mass of the resulting BH, but keeps constant or
even reduces the velocity dispersion of the system \citep{Naab2009}. A
large fraction of brightest cluster galaxies (BCGs) in the local
Universe have relatively low velocity dispersions: \citet{Kluge2023}
measure on average a velocity dispersion of 250 km s$^{-1}$ for their large
sample of BCGs, with only 10\% of objects having $\sigma>300$ km s$^{-1}$.
The $M_{\rm BH}$-$\sigma$ relation translates $\sigma=250$ km s$^{-1}$ into BH
masses around only about $6\expo{8} \solarmass$. Nevertheless, HMBHs are
found in BCGs, the largest ($4\expo{10} \solarmass$) in Holm 15A
\citep{Mehrgan2019}. The most promising way to search for them is to
select massive ETGs, in particular BCGs, with core radii of the order
or larger than 0.6 \kiloparsec\ (Holm 15A has a core radius of 4
\kiloparsec). The Euclid Wide and Deep Surveys will allow us to find
these objects in large numbers and out to redshifts around 1, thanks to
their excellent spatial resolution, large area coverage, and
depth. Here we report the detection of the $1\arcsecf29$ (or 0.45
\kiloparsec) core of NGC 1272, measured on the \Euclid ERO VIS
\citep{Cropper2024} image of the Perseus cluster
\citep{Cuillandre2024,Perseus2024}, and the dynamical determination of
the mass of its BH.

The galaxy is the second brightest elliptical galaxy of Perseus.  With
a total magnitude in the V band corrected for Galactic absorption
$V_{\rm T}^0$ of 11.27 \citep{RC3}, we compute a luminosity of
$L=1.3\expo{11}\solarluminosity$ using 72 Mpc as the distance of the
cluster \citep{Kluge2024}, with which $1^{\prime\prime}$ translates to
0.35 \kiloparsec.  The stellar mass of the galaxy is
$9\expo{11}\solarmass$, using our dynamically determined mass-to-light
ratio of $7\, \solarmass/\solarluminosity$ (see Sect. \ref{Sec_Dyn})
and the effective radius quoted in \citet{RC3} is $57^{\prime\prime}$
or 20 \kiloparsec. With these properties NGC 1272 belongs to the class
of cD galaxies, the most massive ellipticals. We expect such objects
to be triaxial with a low $V/\sigma$ parameter (the ratio between the
mean stellar velocity $V$ and the velocity dispersion $\sigma$ of the
galaxy), to be detected as radio sources and to have extended X-ray
emission \citep{Bender1989}.  Consistent with this, \citet{Veale2017}
present stellar kinematics of NGC 1272 obtained with the VIRUS-P
spectrograph and classify the galaxy kinematically as a
slow-rotator. \citet{Park2017} detect a faint radio source at its
center and \citet{McBride2014} study the properties of the double jets
emerging from the center of the galaxy (which are bent with a
curvature radius of 2 \kiloparsec). \citet{Arakawa2019} detect and
study the X-ray minicorona of the galaxy, measuring a temperature of
0.63 keV and a size of 1.2 \kiloparsec .

The structure of the paper is the following. In Sect. \ref{Sec_Obs} we
describe the photometric and spectroscopic observations of NGC
1272. The dynamical modeling is presented in Sect. \ref{Sec_Dyn}. We draw
our conclusions in Sect. \ref{Sec_Conclusions}, where we discuss the
prospect of exploiting the \Euclid survey \citep{Mellier2024} to find
large cores up to redshift 1 to probe the formation redshift of the
most massive black holes in galaxies.

\section{Observations}
\label{Sec_Obs}

NGC 1272 was observed during the early days of the \Euclid survey, as part
of the pointings covering the Perseus galaxy cluster \citep{Perseus2024},
one of the objects selected for the Early Release Observations (ERO) program.
The VIS \citep{Cropper2024} and NISP \citep{Janke2024} ERO images of
the cluster were reduced as described in \citet{Cuillandre2024}. Based
on this dataset, studies of the Perseus intracluster light and
intracluster globular clusters are described in \citet{Kluge2024} and
of its dwarf galaxy population in \citet{Marleau2024}.

In Sect. \ref{Sec_Photo} we make quantitative
use of the VIS image of NGC 1272, with pixel size and resolution of
0\arcsecf1 and $0\arcsecf17$, respectively. We used the near-infrared images
(with $0\arcsecf3$ pixels) to assess the absence of dust in the
central regions of the galaxy. The complementary spectroscopic
information is described in Sect. \ref{Sec_Spec}.

\subsection{Photometry}
\label{Sec_Photo}

\begin{figure}
\centering
\includegraphics[width=1\columnwidth]{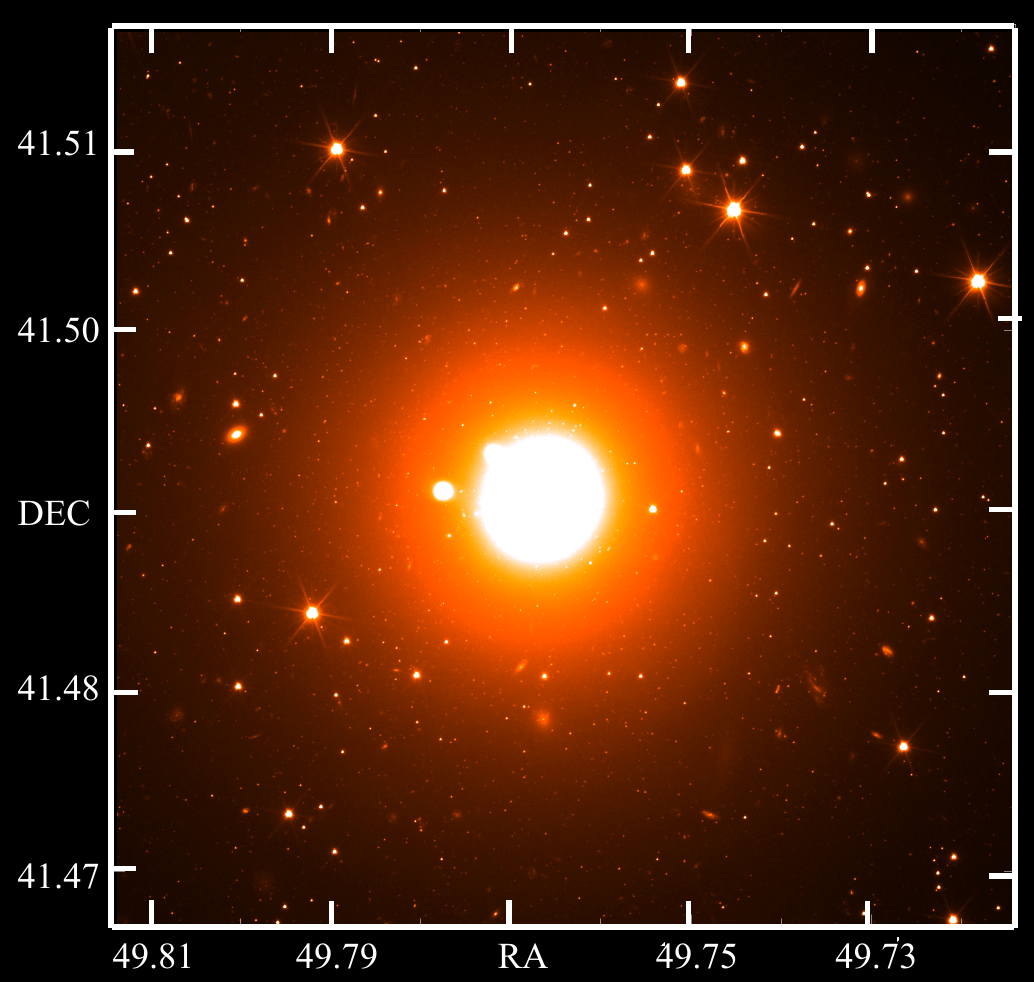} 
\caption{A cutout of the \Euclid VIS image of NGC 1272.}
\label{Fig_imaN1272}
\end{figure}

\begin{figure}
\centering
\includegraphics[width=1\columnwidth]{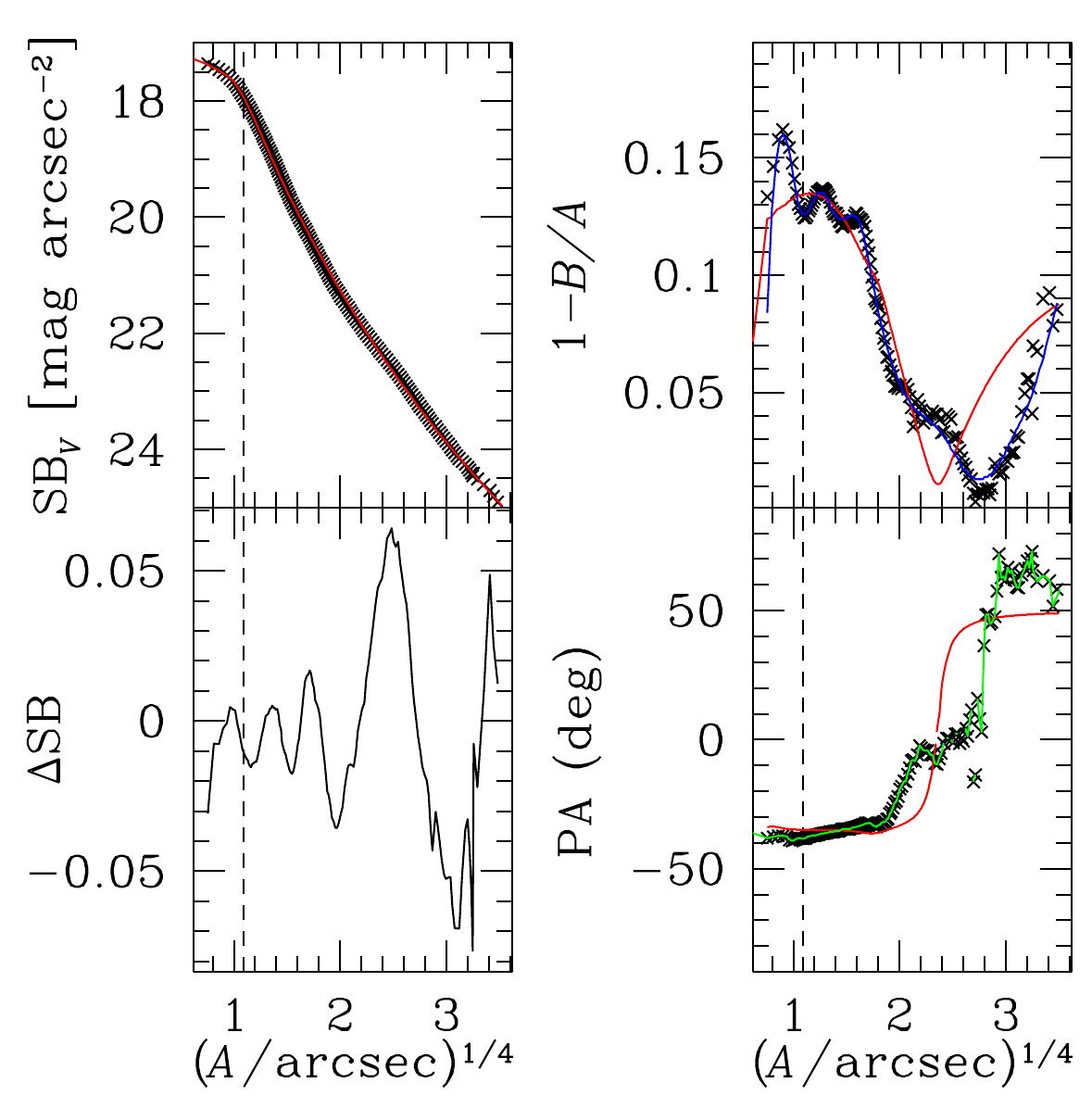} 
\caption{\emph{Top Left:} surface brightness profile of NGC 1272,
  measured from the \Euclid VIS image, calibrated to the $V-$band
  and corrected for Galactic absorption and cosmological dimming,
  as a function of the 1/4 power of the semi-major distance
  $A$ on the sky in arcsec. \emph{Bottom Left:} the difference
  $\Delta${\it SB} between the surface brightness profile of NGC 1272 and the
  surface brightness of the core-S\'ersic+S\'ersic model.
  \emph{Right:} ellipticity $1-B/A$, where $B$ is the semi-minor
  axis length on the sky (top) and PA (bottom) as a function of the 1/4
  power of $A$.  The solid red lines show the core-S\'ersic+S\'ersic
  model. The dashed lines show its core radius. The blue line shows
  the ellipticity profile of the axisymmetric deprojection. The green
  line shows the PA profile of the triaxial deprojection.}
    \label{Fig_photoN1272}
\end{figure}

Figure \ref{Fig_imaN1272} shows a cutout of the \Euclid VIS
image of NGC 1272.  The isophote shape analysis was performed
following \cite{Bender1987}. Figure \ref{Fig_photoN1272} shows the
resulting surface brightness profile calibrated to the $V-$band for
compatibility with the results of \citet{Rusli2013}. We perform the
calibration by integrating the profile in circular apertures, that we
shift to reproduce the aperture photometry listed in
Hyperleda\footnote{\url{http://atlas.obs-hp.fr/hyperleda/}}. Finally,
we adopt the correction for Galactic absorption and cosmological
surface brightness dimming adopted in \citet{RC3} by matching
our aperture magnitude within  $57^{\prime\prime}$, the half-luminosity radius,
to $V_{\rm T}^0+2.5\logten 2$, half the total luminosity of the galaxy.
We measure the
photometry out to $147^{\prime\prime}$ from the center, down to 24.9
mag arcsec$^{-2}$, five times the distance reached by our stellar
kinematics. The galaxy is round, with ellipticities smaller than 0.15
and isophotes showing only small deviations from perfect ellipses. For
radii larger than about 45$^{\prime\prime}$, the center of the
isophotes starts drifting towards the direction of NGC 1275 and the
position angle twists by $70^\circ$. In the inner $1\arcsecf2$ the
surface brightness increase towards the center slows down, pointing to
the presence of a core.

The size of the cores of ETGs has been determined in the past using
the Nuker law \citep{Faber1997} and the core-S\'ersic law
\citep{Graham2003}. Pros and cons of the two approaches have been
discussed at length in the literature and depend on how well the outer
parts of a galaxy can be described by either law. In the following we
rely on both approaches as a way to estimate systematic effects
affecting our measurements.

We start deriving the size of the core of NGC 1272
by fitting a $5000\times5000$ pixels
($250^{\prime\prime}\times 250^{\prime\prime}$) image extracted from
the VIS mosaic with the PSF-convolved core-S\'ersic function provided
by the {\tt Imfit}
code\footnote{\url{https://www.mpe.mpg.de/~erwin/code/imfit/index.html}}
of \citet{Erwin2015}, using the image of a star extracted in the
vicinity of the galaxy as the PSF. Fitting larger cutouts requires
prohibitively large computing time without improving the determination
of the core size. As implemented in  \citet{Erwin2015}, the core-S\'ersic
function is:
\begin{equation}
  \label{eq_coreSersic}
  I_{\rm CS}(r)=I^{\prime}\left[1+\left(\frac{r_{\rm b}}{r}\right)^{\alpha_{\rm CS}}\right]^{\gamma_{\rm CS}/\alpha_{\rm CS}}
  \exp\left[-b_n\left(
  \frac{r^{\alpha_{\rm CS}}+r_{\rm b}^{\alpha_{\rm CS}}}{r_{\rm e}^{\alpha_{\rm CS}}}
      \right)^{1/n\alpha_{\rm CS}}
      \right],
\end{equation}
where
\begin{equation}
\label{eq_Iprime}
I^\prime=I_{\rm CS,b}2^{-\gamma_{\rm CS}/\alpha_{\rm CS}}
\exp\left[b_n\left(2^{1/{\alpha_{\rm CS}}}\frac{r_{\rm b}}{r_{\rm e}}\right)^{1/n}\right]
\end{equation}
and $b_n\sim 2n-1/3+4/405n$. Similarly, the S\'ersic function is:
\begin{equation}
  \label{eq_Sersic}
  I_{\rm S}(r)=I_{\rm e}\exp\left[-b_n \left(\frac{r}{r_{\rm e}}\right)^{1/n}\right].
\end{equation}
Here $r_{\rm e}$ is the half-luminosity radius, $I_{\rm e}$ the
intensity at $r_{\rm e}$, $n$ the S\'ersic index, $r_{\rm b}$ the
break radius and $I_{\rm CS,b}$ the intensity at $r_{\rm b}$,
$-\gamma_{\rm CS}$ is the slope of the power-law inner profile, and
$\alpha_{\rm CS}$ specifies the sharpness of the transition to the
outer, S\'ersic profile. In Table \ref{Tab_CoreSersic} and
\ref{Tab_Sersic} we provide the values of
$\mu_V(r_{\rm e})$ and $\mu_V(r_{\rm b})$ that calibrate the surface
brightness profiles $\mu_{\rm S}=-2.5 \logten
I_{\rm S}/I_{\rm e}+\mu_V(r_{\rm e})$ and
$\mu_{\rm CS}=-2.5\logten I_{\rm CS}/I_{\rm CS,b}+\mu_V(r_{\rm b})$ to the $V-$band.

The core-S\'ersic model reproduces the surface brightness of the
galaxy accurately, with residuals less than 0.1 mag, even if it has a
constant ellipticity and position angle. The resulting parameters of
the fit are given in Table \ref{Tab_CoreSersic}; in particular, the
size of the core is perfectly resolved by the spatial resolution of
the VIS image. According to \citet{Thomas2016}, we expect this to
match the size of the sphere of influence of the central black hole of
the galaxy. The best fitting value of $n$ (21.1) is unrealistically
large, as is the one of $r_{\rm e}$, two orders of magnitudes larger
than the size of the fitted image; this stems from the almost
power-law behavior of the outer profile, typical of BCGs
\citep{Kluge2023}. Both parameters are to be considered as a
convenient parametrization of the galaxy profile out to the limit of
the image and increasing with the image size. More importantly, the
values of $r_{\rm b}$ and $\gamma_{\rm CS}$ do not depend much on this
choice. The statistical errors listed in Table \ref{Tab_CoreSersic}
(and further below in Tables \ref{Tab_Nuker} and \ref{Tab_Sersic}) are
minute, because the number of independent points in the image is
huge. We have rounded them up to the first or second digit.  Fitting
the surface brightness of Fig.\ref{Fig_photoN1272} with a
1-dimensional S\'ersic profile without PSF-convolution delivers
similar results within the systematic errors estimated below.

We explore further the systematic errors affecting the estimation of
the core radius by fitting the same image with the {\tt Imfit}
implementation of the (PSF-convolved) Nuker-law:
\begin{equation}
  \label{eq_Nuker}
  I_{\rm N}(r)=I_{\rm N,b} 2^{(\beta_{\rm N}-\gamma_{\rm N})/\alpha_{\rm N}}
    \left(\frac{r_{\rm b}}{r}\right)^{\gamma_{\rm N}}\left[1+\left(\frac{r}{r_{\rm b}}\right)^{\alpha_{\rm N}} \right]^{(\gamma_{\rm N}-\beta_{\rm N})/\alpha_{\rm N}}.
\end{equation}
Here $-\gamma_{\rm N}$ is the asymptotic logarithmic slope inside
$r_{\rm b}$, $-\beta_{\rm N}$ is the asymptotic outer slope, and the
$\alpha_{\bf N}$ parameter describes the sharpness of the break;
$I_{\rm N,b}$ is the intensity at $r_{\rm b}$ and the surface brightness
profile $\mu_{\rm N}(r)=-2.5\logten I_{\rm N}/I_{\rm N,b}+\mu_V(r_{\rm
  b})$ is calibrated to the V band through the value of $\mu_V(r_{\rm
  b})$ given in Table \ref{Tab_Nuker}. The Nuker fit delivers a
core-size determination similar to what found using the core-S\'ersic
function, see Table \ref{Tab_Nuker}, and, as noted above, describes
reasonably well the outer power-law behavior of the galaxy profile.

\begin{figure}
\centering
\includegraphics[width=1\columnwidth]{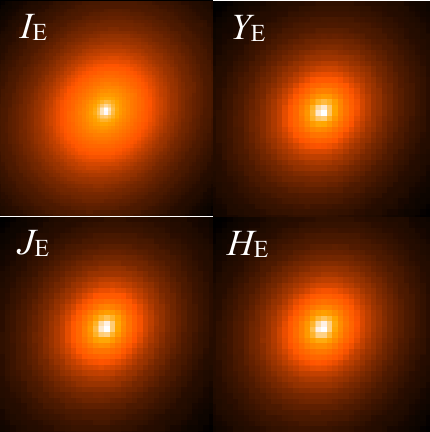} 
\caption{The  \IE, \YE, \JE, and \HE\ cutouts of the inner
  $6^{\prime\prime}\times 6^{\prime\prime}$ of NGC 1272.}
\label{Fig_centerN1272}
\end{figure}

Inspection of the NIR images (see Fig. \ref{Fig_centerN1272})
confirms that the
central region of NGC 1272 is not strongly affected by dust: core
sizes between $1\arcsecf25$ and $1\arcsecf29$ are obtained when
fitting these images. The slope of the surface brightness profile
inside $r_{\rm b}$ is between 0.1 (from the Nuker fit) to 0.2 (from
the core-S\'ersic fit), in the range expected for core ellipticals
\citep{Faber1997}.

\begin{figure}
\centering
\includegraphics[width=1\columnwidth]{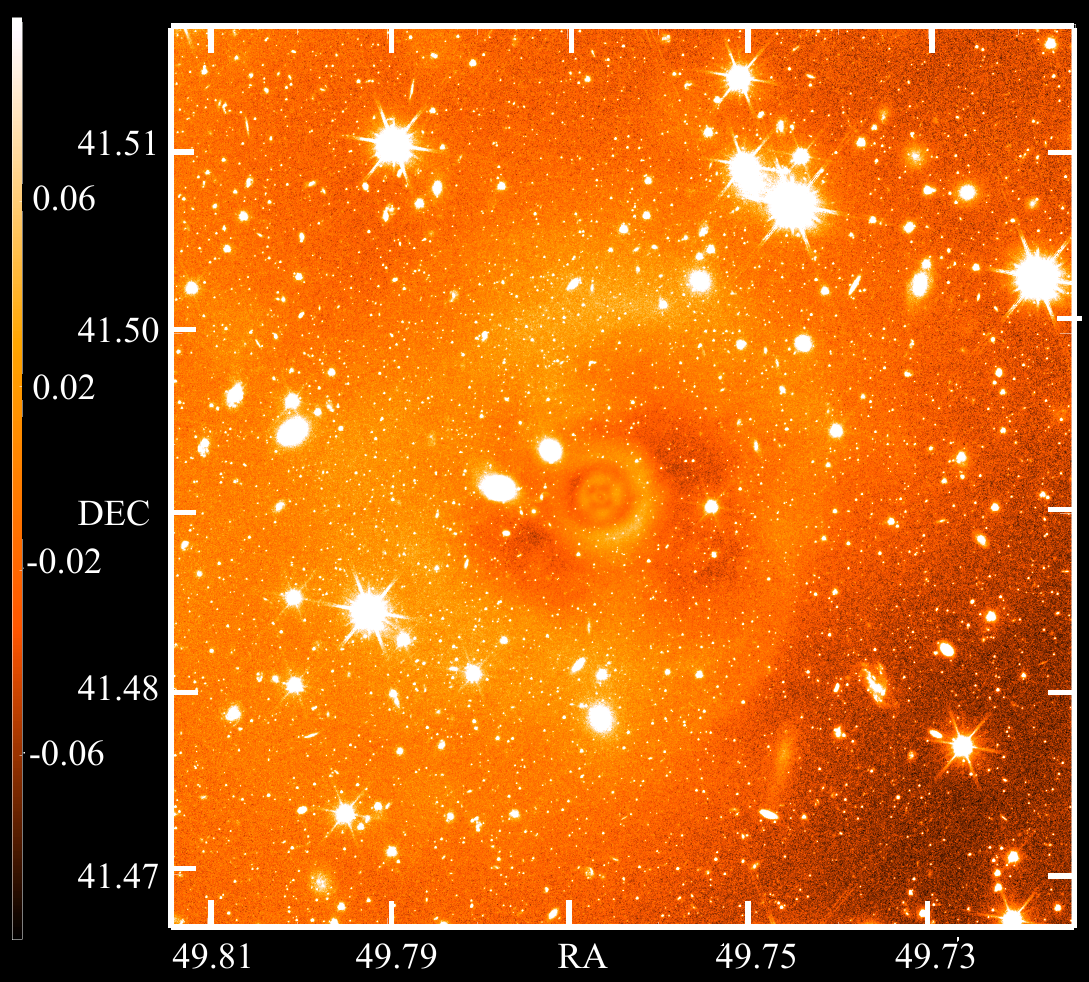} 
\caption{Percentage residuals after subtraction of the core-S\'ersic +
      S\'ersic model.}
\label{Fig_resN1272}
\end{figure}

A more realistic estimate of the effective radius of the galaxy that
catches better the varying ellipticity and PA profiles (see
Fig.~\ref{Fig_photoN1272}) is obtained by fitting a two-component
model, an inner core-S\'ersic plus an outer S\'ersic profile. The
results are listed in Table~\ref{Tab_CoreSersic} and
\ref{Tab_Sersic}. The core radius is somewhat larger and the inner
slope $\gamma_{\rm CS}$ of the profile somewhat shallower than above. We show in
Fig. \ref{Fig_resN1272} the fractional residuals between image
and model that in absolute sense are always smaller than 0.1. The
surface brightness is reproduced with a root mean square (RMS) of 0.028 mag
(see Fig. \ref{Fig_photoN1272}).

\citet{deRijke2009} collected F555W and F814W ACS images of NGC 1272
with the Hubble Space Telescope (HST), with pixel size and resolution
a factor 2 better than our \Euclid VIS images. We perform
core-S\'ersic fits to $179^{\prime\prime}\times 184^{\prime\prime}$
images with {\tt Imfit}, fixing the value of $n$ to the result
obtained from the VIS image. The results are given in Table
\ref{Tab_CoreSersic}, where we calibrate $\mu_{\rm b}=-2.5 \log
I_{\rm b}$ to the $V-$band as above. We measured $r_{\rm b}=1\arcsecf27$ and
$1\arcsecf25$ in the two bands, demonstrating that possible color
gradients do not affect strongly the determination of $r_{\rm b}$.

The statistical errors reported in Table \ref{Tab_CoreSersic},
\ref{Tab_Nuker}, and \ref{Tab_Sersic} are minute due to the large
number of pixels fitted. More significant are the systematic errors
that come from the different fitting functions used to measure $r_{\rm
  b}$. Averaging the five estimates of $r_{\rm b}$ presented above one
gets $1\arcsecf29$, or 0.45 \kiloparsec, with root mean square (RMS)
of $0\arcsecf07$, which we adopt as our measurement error.

\begin{figure}
    \centering
    \includegraphics[width=1\columnwidth]{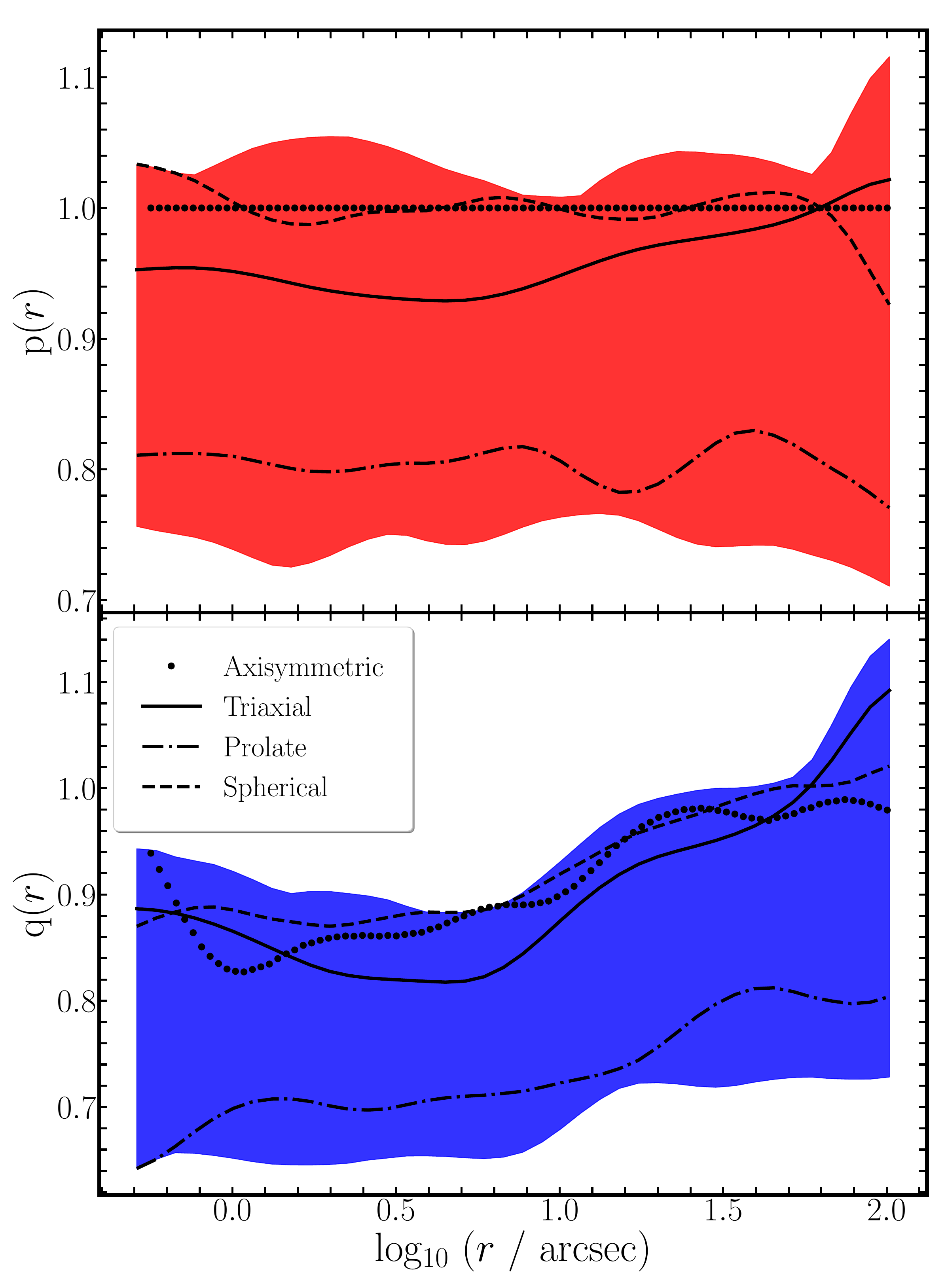} 
    \caption{Profiles of $p(r)$ and $q(r)$ (where $p=b/a$, $q=c/a$ and $a$, $b$ and $c$ are the major,
intermediate and minor semi-axis of the galaxy) derived from the axisymmetric (dotted), triaxial (full line), the spherical (dashed line), and prolate (dashed-dotted line) deprojections
of the galaxy, as a function of the distance $r$ from the center.
The red and blue shaded areas show the whole range of 
      allowed deprojections with ${\rm RMS}\le 1.2\times {\rm RMS_{min}}$ \citep{deNicola2020,deNicola2022a,deNicola2022b}.}
    \label{Fig_pqprofiles}
\end{figure}

We deproject the surface brightness profile using the axisymmetric
deprojection code of \citet{Magorrian1999}, assuming that the galaxy
is edge-on, as is usually done in this case \citep{LipkaThomas2021}.
Other options are explored below, when triaxial
deprojections are considered. The blue line in
Fig. \ref{Fig_photoN1272} shows that this deprojection reproduces the
ellipticity profiles, but cannot reproduce the PA profile (assumed to
be constant in axisymmetric deprojections). The intrinsic flattening profile
$q(r)$, where $q=c/a$ and $a$ and $c$ are the major
and minor semi-axes of the galaxy, derived in this way is
around 0.9 (see Fig. \ref{Fig_pqprofiles}).

We also explore the range of possible triaxial deprojections
following \citet{deNicola2020}. The reconstructed $p$ and $q$ profiles
(where $p=b/a$ and $b$ is the intermediate semi-axis of the galaxy)
are shown in Fig. \ref{Fig_pqprofiles} and demonstrate that the galaxy
is almost spherical and close to axisymmetric, with $p\approx 1$ and
$q\approx 0.9$. For some viewing angles the strong PA radial variation
(see Fig. \ref{Fig_pqprofiles}) forces a twist of the principal axis
with radius, which explains why the $p$ and $q$ profiles can become
larger than one at same distance \citep{deNicola2020}.

The deprojection with the lowest RMS
in surface brightness is obtained at angles
$(\theta,\phi,\psi)=(64^\circ,124^\circ,23^\circ)$, roughly $26^\circ$
above the equatorial plane and about $34^\circ$ away from the
intermediate axis. However, it is clear that reconstructing the true
orientation of the galaxy is almost impossible, given its almost
spherical geometry. The green line in Fig. \ref{Fig_photoN1272} shows
that this deprojection does reproduce the PA profile (additionally to
the ellipticity profile). We further explore alternative deprojections
with comparably good surface brightness RMS: a (mildly) prolate and an
(almost) spherical deprojection (similar to the axisymmetric one, but
matching the PA twist).  Both are obtained assuming that the
line of sight is along the major axis of the galaxy and starting the
deprojection routine with constant profiles $q(r)=0.7$ and $q(r)=0.95$
in the prolate and spherical cases, respectively.

\begin{table*}
    \centering
    \caption{Parameters of the core-S\'ersic best fits.}
    \begin{tabular}{ccccccccc}
    \hline
    \hline
\noalign{\smallskip}
    Image &    PA & $1-B/A$ & $n$ & $r_{\rm e} $ & $\mu_V(r_{\rm b})$ & $r_{\rm b}$ & $\alpha_{\rm CS}$ & $\gamma_{\rm CS}$ \\
 & [deg] &    &     & $[^{\prime\prime}] $ &  [mag arcsec$^{-2}$] & $[^{\prime\prime}]$ & & \\
\noalign{\smallskip}
    \hline
\noalign{\smallskip}
      VIS & $-34.1\pm0.1$ & $0.10\pm 0.01$ & $21.14\pm 0.02$ & {\it 22807}$^a\pm$ {\it 126} & $17.83\pm 0.01$ & $1.24\pm0.01$ & $3.99\pm 0.01$ & $0.21\pm0.01$ \\  
VIS$^b$ &  $-35.4\pm0.1$ & $0.15\pm 0.01$ & $12.1\pm 0.1$ & $34.9\pm 1.0$ & $17.93\pm 0.01$ & $1.41\pm0.01$ & $2.47\pm 0.01$ & $0.15\pm0.01$ \\ 
F555W & $-35.1\pm0.1$ & $0.11\pm 0.01$ & 21.14 & {\it 18392}$^a\pm$ {\it 4} & $17.85\pm 0.01$& $1.27\pm0.01$ &  $3.41\pm 0.01$ &  $0.17\pm0.01$ \\
F814W & $-34.8\pm0.1$ & $0.11\pm 0.01$ & 21.14 & {\it 16246}$^a\pm$ {\it 1} & $17.83\pm 0.01$& $1.25\pm0.01$ &  $3.64\pm 0.01$ &  $0.19\pm0.01$ \\
\noalign{\smallskip}
\hline
\noalign{\smallskip}
    \end{tabular}
    \tablefoot{We list the fitted image (column 1), the position angle
      (column 2), the ellipticity (column 3), the values of
      $n$, $r_{\rm e}$, $\mu_v(r_{\rm b})$, $r_{\rm b}$, $\alpha_{\rm CS}$,
      and $\gamma_{\rm CS}$ (columns 4, 5, 6, 7, and 8, respectively),
      see Eq. \ref{eq_coreSersic}.\\
    \tablefoottext{a}{The value is unrealistically large, see text.}
    \tablefoottext{b}{With second S\'ersic component, see Table \ref{Tab_Sersic}.}}
    \label{Tab_CoreSersic}
\end{table*}

\begin{table*}
    \centering
    \caption{Parameters of the Nuker best fit.}
    \begin{tabular}{cccccccc}
    \hline
    \hline
\noalign{\smallskip}
    Image &   PA & $1-B/A$ & $\mu_V(r_{\rm b})$ & $r_{\rm b}$ & $\alpha_{\rm N}$ & $\beta_{\rm N}$ & $\gamma_{\rm N}$ \\   
 & [deg] &    &  [mag arcsec$^{-2}$] & $[^{\prime\prime}]$ & & &\\
\noalign{\smallskip}
      \hline
\noalign{\smallskip}
 VIS &$-33.9\pm0.1$ & 0.1 & 17.86 & $1.29\pm0.01$ & $2.35\pm 0.01$& $1.42\pm 0.01$& $0.12\pm0.01$ \\  
       \hline
    \end{tabular}
    \tablefoot{We list the fitted image (column 1), the position angle (column 2), the ellipticity (column 3), the values of $\mu_v(r_{\rm b})$, $r_{\rm b}$, $\alpha_{\rm N}$, $\beta_{\rm N}$, and $\gamma_{\rm N}$ (columns 4, 5, 6, 7, and 8, respectively), see Eq. \ref{eq_Nuker}.}
    \label{Tab_Nuker}
\end{table*}
\begin{table*}
    \centering
    \caption{Parameters of the second S\'ersic component.}
    \begin{tabular}{cccccc}
    \hline
    \hline
\noalign{\smallskip}
    Image &   PA & $1-B/A$ & $n$ & $r_{\rm e}$ & $\mu_V(r_{\rm e})$ \\
          & [deg] &    &     & $[^{\prime\prime}]$ & [mag arcsec$^{-2}$] \\
\noalign{\smallskip}
    \hline
\noalign{\smallskip}
    VIS   &   $50.3\pm0.1$ & $0.14\pm0.1$ & $2.62\pm0.01$ & $140.8\pm0.1$ & $25.04\pm0.01$\\
\noalign{\smallskip}
       \hline
    \end{tabular}
    \tablefoot{We list the fitted image (column 1), the position angle (column 2), the ellipticity (column 3), the values of
    $n$, $r_{\rm e}$, $\mu_v(r_{\rm e})$ (columns 4, 5, and 6, respectively), see Eq. \ref{eq_Sersic}. }
    \label{Tab_Sersic}
\end{table*}

\subsection{Spectroscopic observations and kinematics}
\label{Sec_Spec}

We observed NGC 1272 spectroscopically with the Visible Integral-field
Replicable Unit Spectrograph (VIRUS) at the Hobby-Eberly Telescope
(HET) on 3 March 2022. The pointing of the telescope was optimized to
observe NGC 1275; as a result the integral field unit (IFU) covering
NGC 1272 was slightly off-center and did not uniformly cover the
galaxy. The seeing reported during the observations was ${\rm
  FWHM}=2\arcsecf36$. The diameter of the single fiber is
$1\arcsecf5$.  Given the size of the core measured above (a diameter
of $2\arcsecf6$), the spatial resolution of this data set is (just)
enough to resolve the sphere of influence of the central black hole of
the galaxy. \citet{Rusli2013b} find that in such a case an unbiased
recovery of the BH mass is possible if the dark matter halo of the
galaxy is taken into account in the dynamical modeling, as we are
doing here, see below. The data cover a wavelength interval ranging
from 3470 \AA\ to 5540 \AA\ with a spectral resolution of $5.6$ \AA.

\begin{figure*}
    \centering
    \includegraphics[width=2\columnwidth]{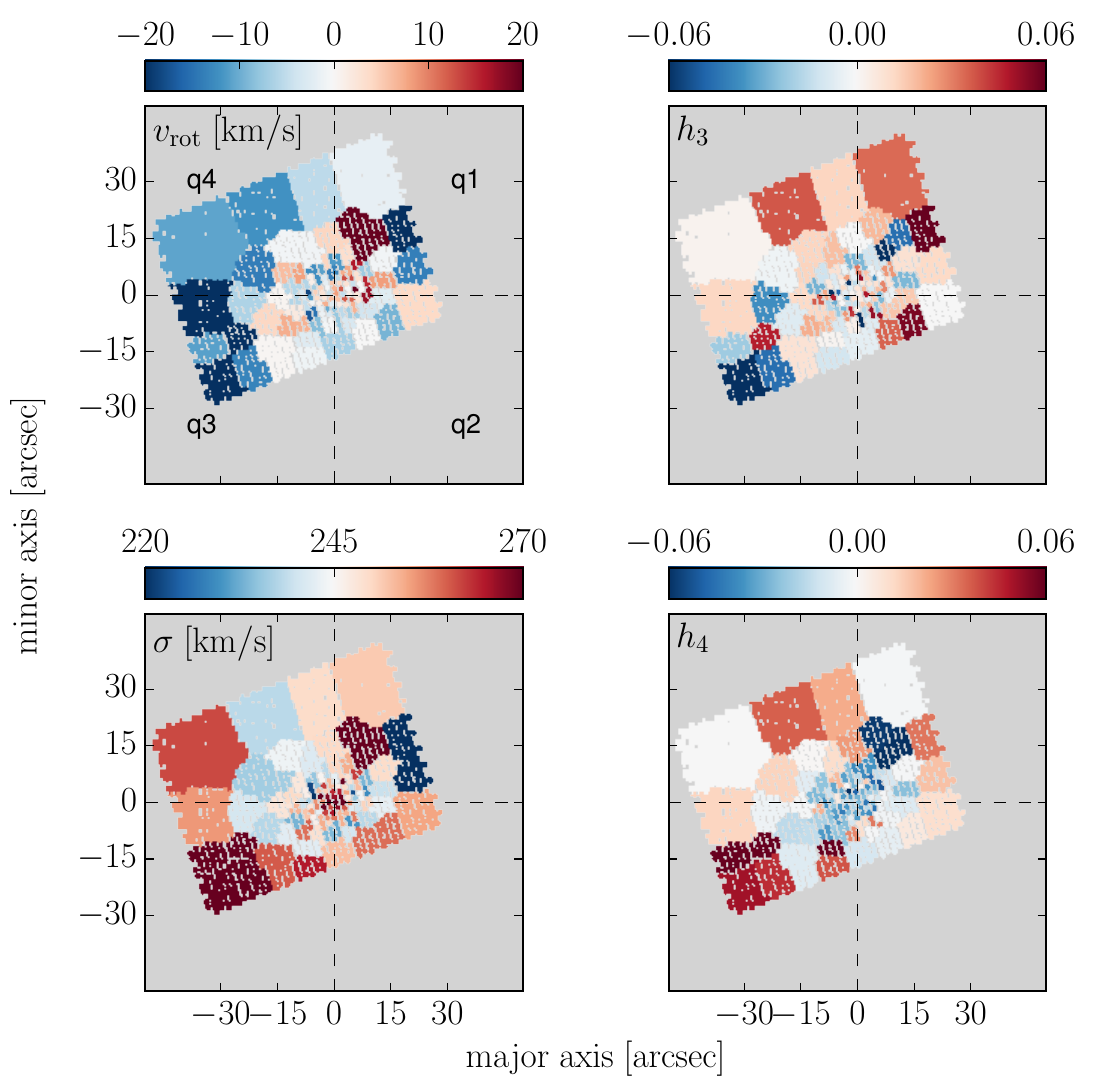} 
    \caption{Two-dimensional stellar kinematics of NGC 1272.
      The horizontal and vertical dashed lines show the major and minor
      axes of the galaxy, respectively. North is up and east is to the left.}
    \label{Fig_2dimKinN1272}
\end{figure*}

\begin{figure*}
    \centering
    \includegraphics[width=2\columnwidth]{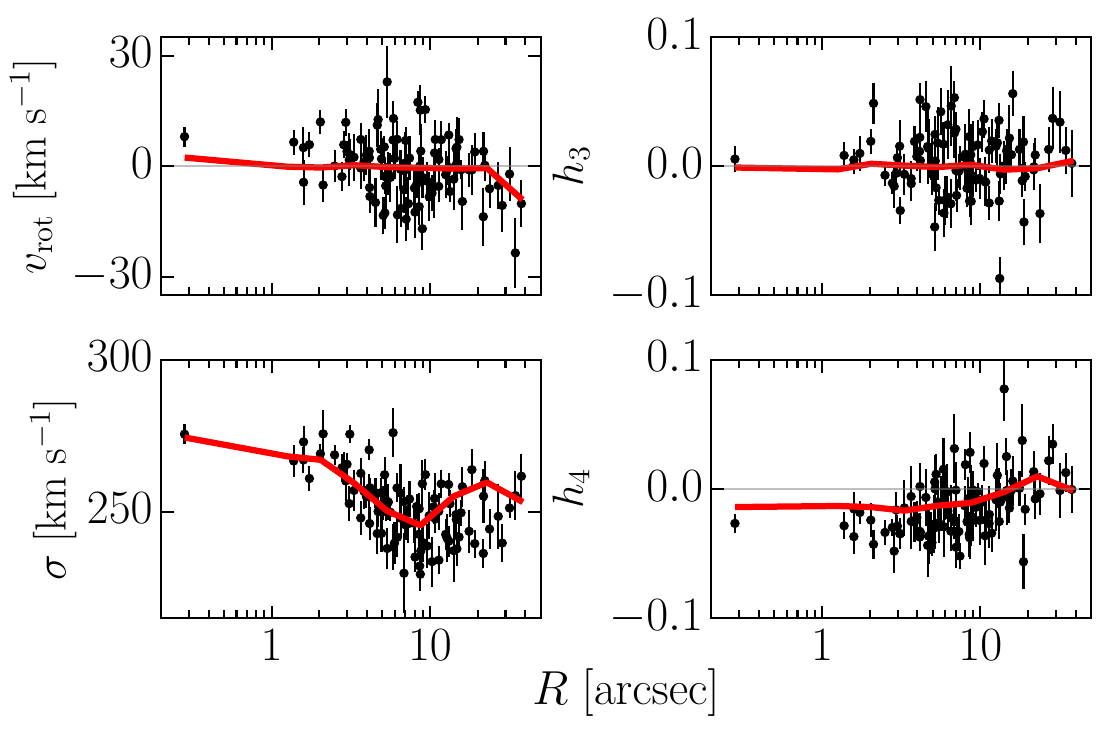} 
    \caption{Radial stellar kinematics of NGC 1272. The red lines show the axisymmetric fit to the stellar kinematics of the galaxy (black data points with error bars) as a function of the distance $R$ from the center of the galaxy on the sky.}
    \label{Fig_radialKinN1272}
\end{figure*}

We used the Voronoi tessellation method of \citet{Cappellari2003} to
spatially bin the spectral data for a target average signal-to-noise
(S/N) of 40. With this target S/N, spectra were only binned together
starting approximately $3^{\prime\prime}$ from the center of the
galaxy, thus maximizing the spatial resolution of our data within the
core region. In this way we ended up with a total of 110 spatial bins.
We measured the stellar kinematics using {\tt WINGFIT} (Thomas, in
prep.), which delivers optimally smoothed non-parametric line-of-sight
velocity distributions (LOSVDs) using the model optimization approach
of \citet{ThomasLipka2022}. The stellar kinematic fits were performed
using the MILES library \citep{Sanchez2006} of stellar
templates. Following the strategy laid out in \citet{Mehrgan2023}, we
performed a careful pre-selection of templates in order to minimize
distortions of the LOSVDs due to template mismatch.  To this end we
fitted the average spectrum of the central $2^{\prime\prime}$ of the
galaxy using all the templates of the MILES library with a
Gauss-Hermite LOSVD that was fixed to be symmetric around a
line-of-sight velocity of zero. We then used the 18 templates that in
the best fit received a non-zero weight as the template set with which
we subsequently fitted all bins of the galaxy with non-parametric
LOSVDs (without the symmetry constraint on the LOSVD).  Also following
\citet{Mehrgan2023}, we used no additive polynomials in the fit and
only a minimal third-order multiplicative polynomial.  Fits were
performed in the wavelength interval between 4700 and 5400 \AA.  The
resulting 2-dimensional kinematic maps are shown in
Fig. \ref{Fig_2dimKinN1272}; the corresponding radial profiles can be
seen in Fig. \ref{Fig_radialKinN1272}. We measured the stellar
kinematics out to a maximum distance of $38^{\prime\prime}$, or 0.66
times the effective radius quoted by \citet{RC3}. The galaxy has small
mean rotation (at most $v=20$ km s$^{-1}$), a relatively low velocity
dispersion $\sigma$ around 250 km s$^{-1}$, increasing to 270 km
s$^{-1}$ towards the center (admittedly with only one point within 1
arcsec from the center), almost zero third-order Hermite parameter
$h_3$ and zero fourth-order Hermite parameter $h_4$, decreasing to
about $-0.05$ towards the center. The data presented by
\citet{Veale2017} match these findings, though their $h_4$ is always
about $0$.  We averaged $(v^2+\sigma^2)^{0.5}$, with equal or
luminosity weights, to get an estimate of $\sigma_{\rm e}$, the
velocity dispersion within the half-luminosity radius (even if our
stellar kinematics reaches out only two-thirds of $r_{\rm e}$, see
above). We find $\sigma_{\rm e}=247 \pm 3$ km s$^{-1}$, which we adopt
in Sect. \ref{Sec_Conclusions}.

\begin{figure}
    \centering
    \includegraphics[width=1\columnwidth]{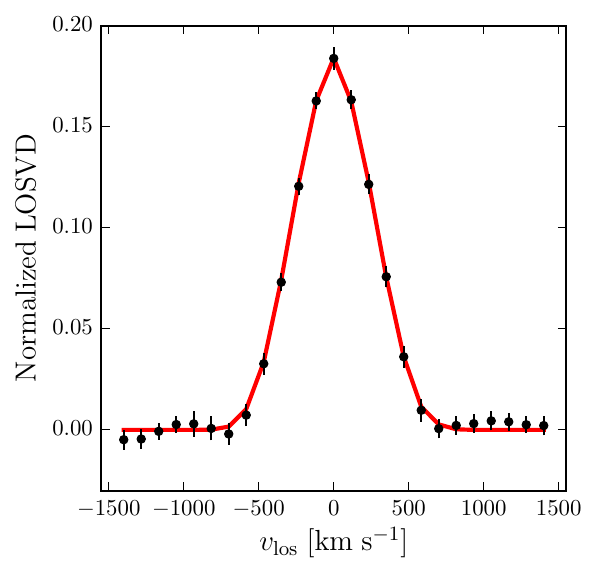} 
    \caption{Line-of-sight velocity distribution measured at 4 arcsec from the center of NGC 1272 (filled circles with error bars). The red line connects the values provided by the base model.}
    \label{Fig_LOSVD}
\end{figure}

For the subsequent dynamical analysis we sampled and fit the non-parametric
LOSVDs between $\pm 1400\,$km s$^{-1}$, with $N_{\mathrm{vel}} = 25$
velocity bins, not the Hermite parameters.
An example LOSVD measured near the center of the galaxy
is shown in Fig.~\ref{Fig_LOSVD}, where the red line connects the values
produced by the best-fitting base model. Finally, to ensure that we
are able to estimate the uncertainties of our dynamical models we
split our data into four quadrants (indicated by q1, q2, q3, and q4 in
Fig. \ref{Fig_2dimKinN1272}) along the minor and major axes of the
galaxy for the axisymmetric dynamical models, and into two halves
(quadrants q1/q4, northern, and q2/q3, southern) split by the major
axis for the triaxial analysis. By modeling quadrants/halves
independently, we can estimate the uncertainties of our best-fit
modeling parameters from the scatter between them. However, the
orientation and positioning of the IFU give us a much better coverage
of the q3 and q4 quadrants, or the east side of the galaxy. Therefore,
we expect the most reliable dynamical constraints to come from these
regions.

\section{Dynamical modeling}
\label{Sec_Dyn}

Given the results presented in Fig. \ref{Fig_pqprofiles} (NGC 1272 is
almost spherical and axisymmetric, but triaxiality and a prolate shape
cannot be excluded), we construct both axisymmetric and
triaxial Schwarzschild models of the galaxy.

The axisymmetric modeling is similar to that of \citet{Mehrgan2024},
with the following modifications. We fit the four quadrants both
independently and together, determining the mass of the central black
hole $M_{\rm BH}$ and the stellar mass-to-light ratio $\Upsilon_\ast$
(i.e., no radial variations of $\Upsilon_\ast$ are considered, because
of the rather coarse and sparse sampling of our stellar
kinematics). We use a spherical \citet{Zhao1996} halo with $\alpha=1$
and $\beta=3$, defined by $\rho_{10}$, the dark matter (DM) density at
10 \kiloparsec, $r_{\rm s}$, the scale radius of the halo
allowed to vary up to the largest distance probed by our
  kinematics \citep{Lipka2024}, and $\gamma_{\rm DM}\ge0$, the inner
slope of the DM halo:

\begin{equation}
\label{eq_zhao}
\rho_{\rm DM}(r)=\frac{k}{
\left( r/r_{\rm s} \right)^{\gamma_{\rm DM}}
{\left(1+r/r_{\rm s}\right)^{3-\gamma_{\rm DM}}}
},
\end{equation}
and $k=\rho_{10}\left(10\,{\rm kpc}/r_{\rm s}\right)^{\gamma_{\rm DM}}\left(1+10\,{\rm kpc}/r_{\rm s}\right)^{3-\gamma_{\rm DM}}$.

The triaxial modeling
follows \citet{deNicola2024} and uses the Schwarzschild code {\tt SMART}
\citep{Neureiter2021} to determine $M_{\rm BH}$ and $\Upsilon_\ast$,
considering a DM halo that is triaxial, described by its shape
parameters $p_{\rm DM}$, $q_{\rm DM}$ plus $\rho_{10}$ and $\gamma_{\rm DM}$, fixing
$r_{\rm s}$ to a large value (158 \kiloparsec).
We model the northern (quadrants q1 and
q4 in Fig. \ref{Fig_2dimKinN1272}) and southern (quadrants q2 and q3 in
Fig. \ref{Fig_2dimKinN1272}) halves of the galaxy separately to assess
the systematic uncertainties.

In both the axisymmetric and triaxial cases we maximize the quantity
$\hat{S}=S-\hat{\alpha} \, \chi^2$ to determine the orbital weights.
Here $\chi^2$ is calculated from the model fit to the observed
non-parametric LOSVDs, and $S$ is the Boltzmann entropy
\citep{Thomas2004}. The deprojected light distributions are used as a
constraint and the parameter $\hat{\alpha}$ is the smoothing of the
models, determined following the prescriptions of
\citet{LipkaThomas2021} and \citet{ThomasLipka2022}, which involve the
determination of the effective degrees of freedom $m_{\rm eff}$. The
parameters $M_{\rm BH}$, $\Upsilon_\ast$, $\rho_{10}$, and $r_{\rm s}$
(in both the axisymmetric and trixial cases), plus $p_{\rm DM}$ and
$q_{\rm DM}$ in the triaxial case, are determined by minimizing the
generalized Akaike information criterion ${\rm AIC}_{\rm p}=\chi^2+2\,
m_{\rm eff}$ over a grid of $\hat{\alpha}$ values.

\begin{figure}
    \centering
    \includegraphics[width=\columnwidth]{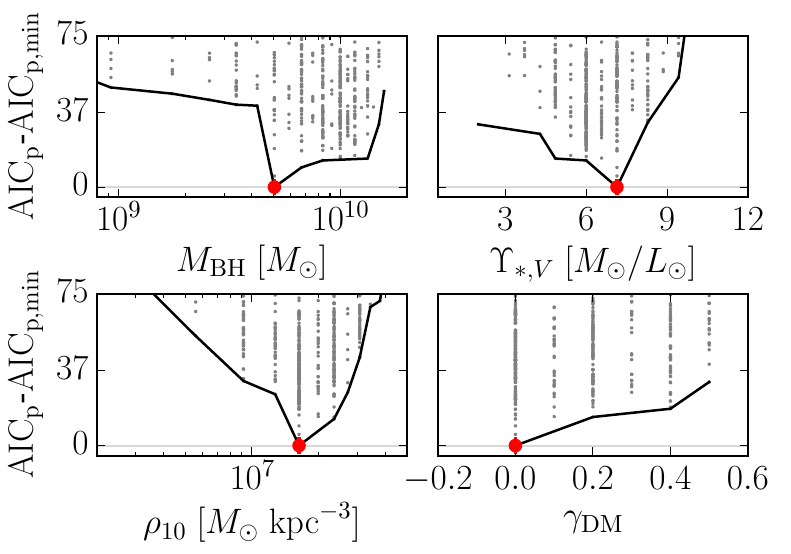} 
    \caption{The results of the axisymmetric modeling of NGC 1272 of the complete stellar kinematic dataset.  As a
      function of the quality of the fits measured by the ${\rm AIC}_{\rm p}$ we show:
      from left to right, from top to bottom: the black hole mass $M_{\rm BH}$; 
      the dynamical
      $V-$band mass-to-light ratio $\Upsilon_{\ast,V}$; the DM density at 10 \kiloparsec\ $\rho_{10}$; the inner slope of the DM density profile $\gamma_{\rm DM}$.
    The gray points show the individual models, the red dots show the best-fitting model, the black lines the lower envelope of the gray points distributions. }
    \label{Fig_axiModelN1272}
\end{figure}

The resulting axisymmetric best-fits to the kinematics are shown in
Fig. \ref{Fig_radialKinN1272}; the derived parameters for the
different fit types are listed in Table \ref{Tab_DynMod}. Our base
result is the axisymmetric model of the entire stellar kinematic data
set. It fits the kinematics very well (see red line in
Fig. \ref{Fig_radialKinN1272}), delivering a reduced $\chi^2$ of
$\chi^2/(N_{\rm data}-m_{\rm eff})=0.91$. Figure
\ref{Fig_axiModelN1272} shows $M_{\rm BH}$, $\Upsilon_{\ast, V}$,
$\rho_{10}$, and $\gamma_{\rm DM}$ as a function of the quality of the
fit measured by the ${\rm AIC}_{\rm p}$ value. Every parameter is well
constrained, with small statistical errors (so small that we do not
quote them in Table \ref{Tab_DynMod}). In particular, we detect a
black hole of $5\expo{9} \solarmass$, a mass-to-light ratio
$\Upsilon_{\ast,V}$ of $7.1\, \solarmass/\solarluminosity$, in between
the values derived from our stellar population analysis for the Kroupa
or the Salpeter initial mass function (IMF, see below), a DM density
at 10 \kiloparsec\ (approximately $30^{\prime\prime}$) similar to the
values reported for other massive elliptical galaxies
\citep{Mehrgan2024} and a cored DM density profile.  We compute the
radius $r_{\rm SOI}$ of the sphere of influence of the black hole as
the distance from the center where the total mass (stellar plus DM
without black hole)
equals $M_{\rm BH}$ \citep{Thomas2016}.  Our $r_{\rm SOI}$ matches the
value of $r_{\rm b}$ and is larger than half the FWHM of the seeing of
the spectroscopic observations. This, together with the modeling of
the dark matter halo of the galaxy, allows an unbiased estimate of the
BH mass \citep{Rusli2013b}.

We gauge our (systematic) errors by looking first at the axisymmetric
modeling of the two quadrants covering a decent part of the galaxy, q3
and q4. Here the BH mass can be as low as $1.8\expo{9} \solarmass$ and
$\Upsilon_{\ast,V}$ as large as 8.9 $\solarmass/\solarluminosity$,
with slightly larger DM densities. Further insights into our
systematic errors are gained from the triaxial {\tt SMART} modeling. All
models fit the kinematic data well, with $\chi^2/(N_{\rm data}-m_{\rm
  eff})$ between 0.6 and 1.0 The best-fitting triaxial model delivers
$M_{\rm BH}=(5.9\pm 1.7)\expo{9} \solarmass$, averaging over the two
halves of the galaxy; the smallest and largest values for the BH mass
are obtained fitting the southern half of the galaxy (where the
kinematic coverage is relatively sparse) in the prolate and spherical
cases, respectively.  The dynamical stellar mass-to-light ratio
$\Upsilon_{\ast,V}$ ranges from 4 to 7.7
$\solarmass/\solarluminosity$.  The density of the DM halo ( $\logten
\rho_{10}/[M_\odot\,\kiloparsec^{-3}] = 7.3\pm 0.3$, averaging over all
triaxial models and halves) agrees with the axisymmetric result. The
slope $\gamma_{\rm DM}$ of the DM density profile is smaller than 1,
but possibly steeper than the cored halo determined
axisymmetrically. In Table \ref{Tab_DynMod} we quote as errors of our
base result the RMS of the nine listed best-fitting models.
Finally, the best fitting
shape of the DM halo is spherical ($p_{\rm DM}=q_{\rm DM}=1$),
with only the prolate case delivering $p_{\rm DM}=0.9$.

\begin{figure}
    \centering
    \includegraphics[width=1\columnwidth]{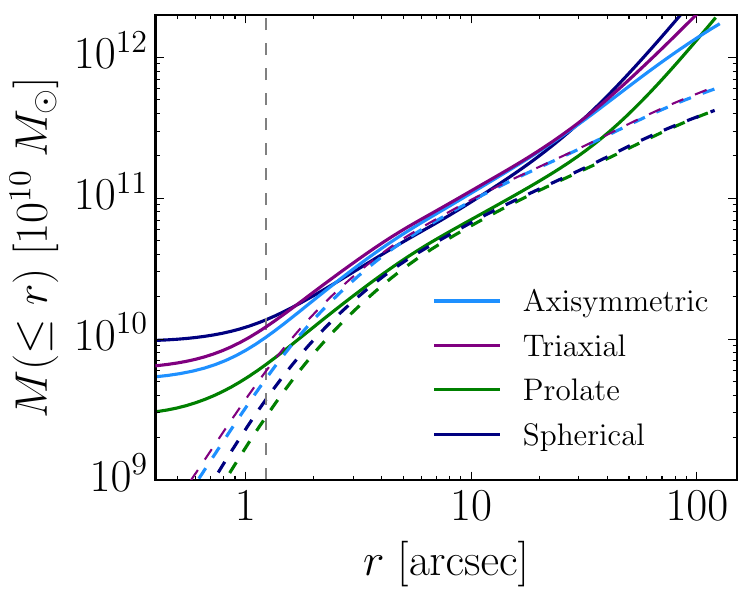} 
    \caption{Spherically-averaged mass profiles of NGC 1272. The solid and dashed lines show the total and stellar profiles, respectively. The triaxial,
      prolate, and spherical models are averaged over the two sides. The vertical dashed line shows the position of the core radius.}
    \label{Fig_massprofile}
\end{figure}

Figure \ref{Fig_massprofile} summarizes the spherically averaged
stellar and total mass profiles derived by the dynamical models we
considered. The total mass distribution is robustly determined in the
region probed by the measured kinematics, with only small deviations
between the different models. Inside the sphere of influence (roughly
the size of the core) the differences between the profiles reflect the
observed scatter in the black hole mass. The stellar mass profiles
scale according to the derived $\Upsilon_{\ast,V}$ values. The dark
halo mass equals the stellar mass at approximately the outermost
radius probed by our stellar kinematics. Using the best-fit value
$\Upsilon_{\ast,V}=7.1 \, \solarmass/\solarluminosity$, we estimate
the total stellar mass of the galaxy from the total luminosity
$L=1.3\expo{11}\solarluminosity$ quoted in the Introduction to be
$9\expo{11}\solarmass$, which we use in Sect. \ref{Sec_Conclusions}.

\begin{figure}
    \centering
    \includegraphics[width=1\columnwidth]{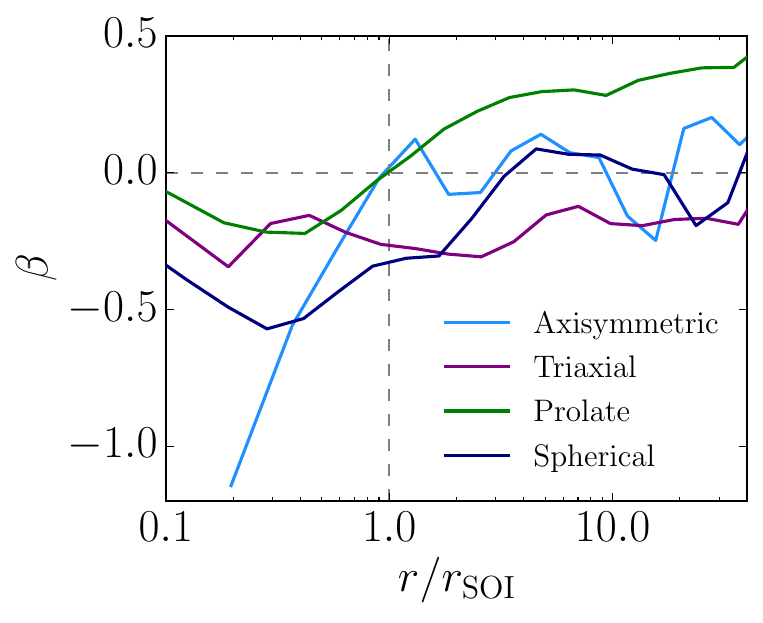} 
    \caption{Anisotropy profiles $\beta$  (where $\beta=1-\sigma_{\rm
  T}^2/\sigma_{\rm R}^2$, and $\sigma_{\rm T}$ and $\sigma_{\rm R}$ are
the spherical tangential and radial velocity dispersions, respectively) of the different models. The triaxial,
      prolate and spherical models are averaged over the two sides. The distances to the center are in units of the core radius.}
    \label{Fig_anisotropy}
\end{figure}

The anisotropy $\beta$ profile (where $\beta=1-\sigma_{\rm
  T}^2/\sigma_{\rm R}^2$, and $\sigma_{\rm T}$ and $\sigma_{\rm R}$ are
the spherical tangential and radial velocity dispersions, respectively)
is not particularly well constrained, but displays the typical feature
of core ellipticals \citep{Thomas2014}. Figure \ref{Fig_anisotropy}
shows that the $\beta$ profile of the base model becomes tangentially
anisotropic within the core radius (the result of core scouring) and 
more isotropic in the outer part, similarly to the spherical model. The
triaxial model is overall mildly tangentially anisotropic, while the prolate
model is radially anisotropic outside the core.

Measuring Lick indices and fitting them with the simple stellar
population models of \citet{Thomas2003} and \citet{Maraston2005}, we
find that the best-fit has a simple stellar population as old as the
Universe, a slightly above Solar metallicity and is more than a factor
of two overabundant in $\alpha$-elements. The derived $V$-band
mass-to-light ratio is $6\,\solarmass/\solarluminosity$ with a
Kroupa-IMF and $8\,\solarmass/\solarluminosity$ with a Salpeter IMF.
This matches the dynamically
determined $\Upsilon_{\ast,V}$, without unambiguously preferring
one of the two options.

Finally, we estimate the mass that has been expelled from the core
during its formation. We consider the core-S\'ersic solution obtained
with the second S\'ersic component (see second line of Table
\ref{Tab_CoreSersic}) and consider the S\'ersic function with $n=12.1$,
$r_{\rm e}=34\arcsecf9$ and $\mu_V(r_{\rm e})=23.57$ that reproduces
the core-S\'ersic solution outside the core region. We integrate the
luminosity difference, or the luminosity deficit $L_{\rm def}$,
between the two functions out to $8^{\prime\prime}$, finding $L_{\rm
  def}=2.7\times10^9\solarluminosity$. Using the dynamically
determined $\Upsilon_{\ast,V}=7\,\solarmass/\solarluminosity$, this
translates into a mass deficit of $M_{\rm def}=1.9\times
10^{10}\solarmass$, or $3.8\, M_{\rm BH}$, in the range found by
\citet{Rusli2013}. According to the simulations of
\citet{GualandrisMerritt2008}, mass deficits up to $5\times M_{\rm
  BH}$ can result from single dry mergers.

\section{Conclusions}
\label{Sec_Conclusions}

\begin{figure*}[ht!]
    \centering
    \includegraphics[width=.67\columnwidth]{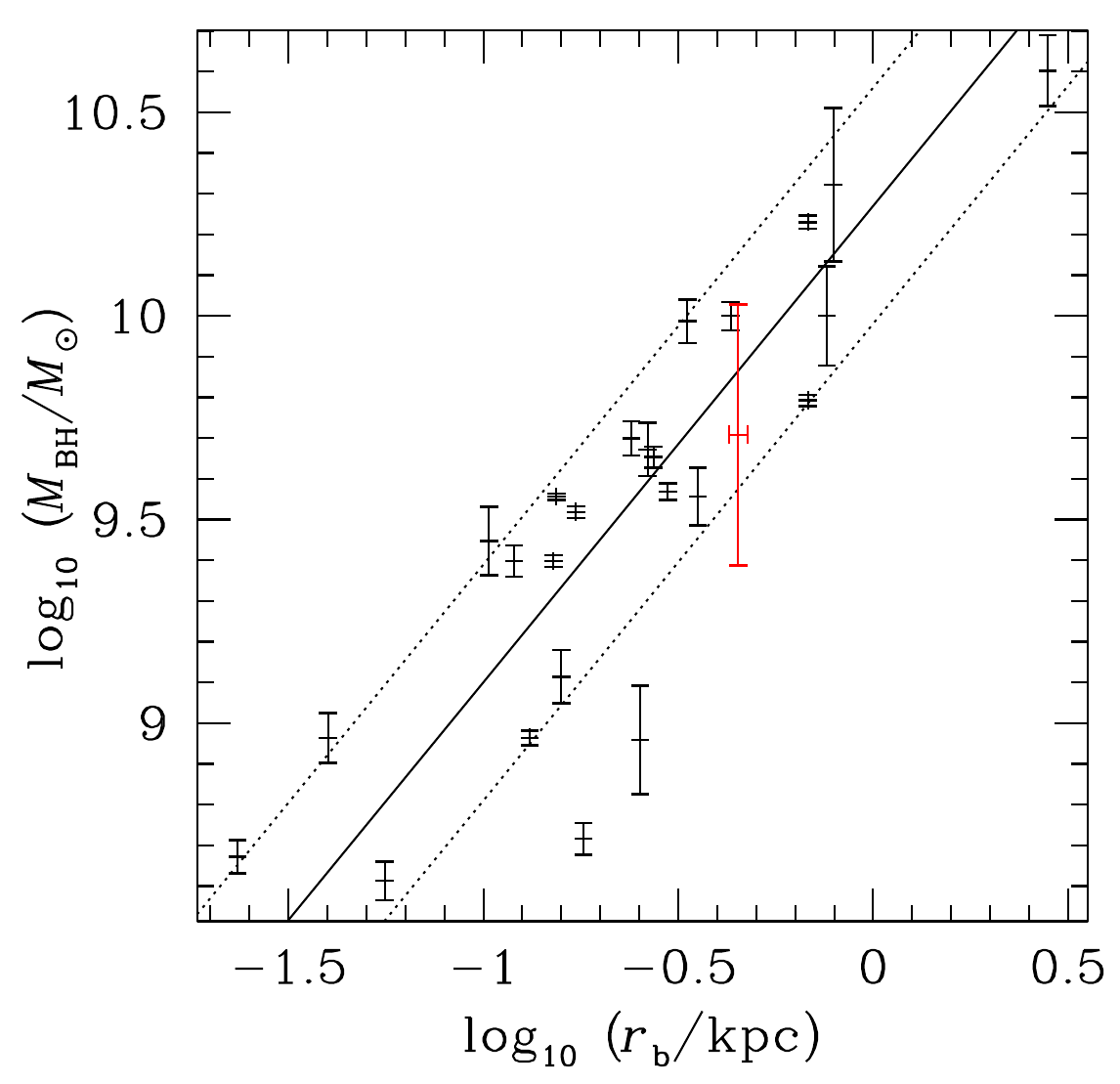} 
    \includegraphics[width=.67\columnwidth]{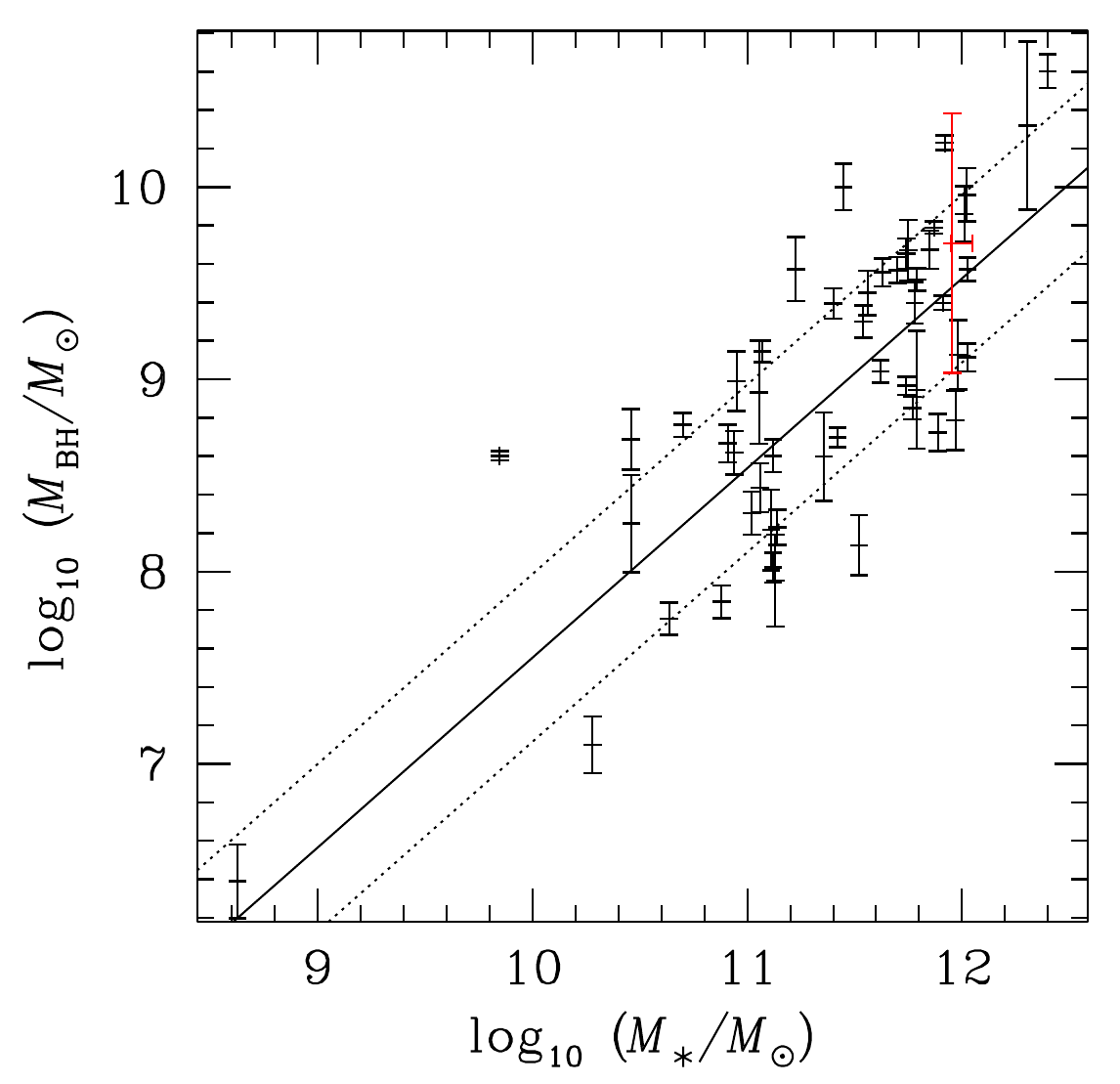} 
    \includegraphics[width=.67\columnwidth]{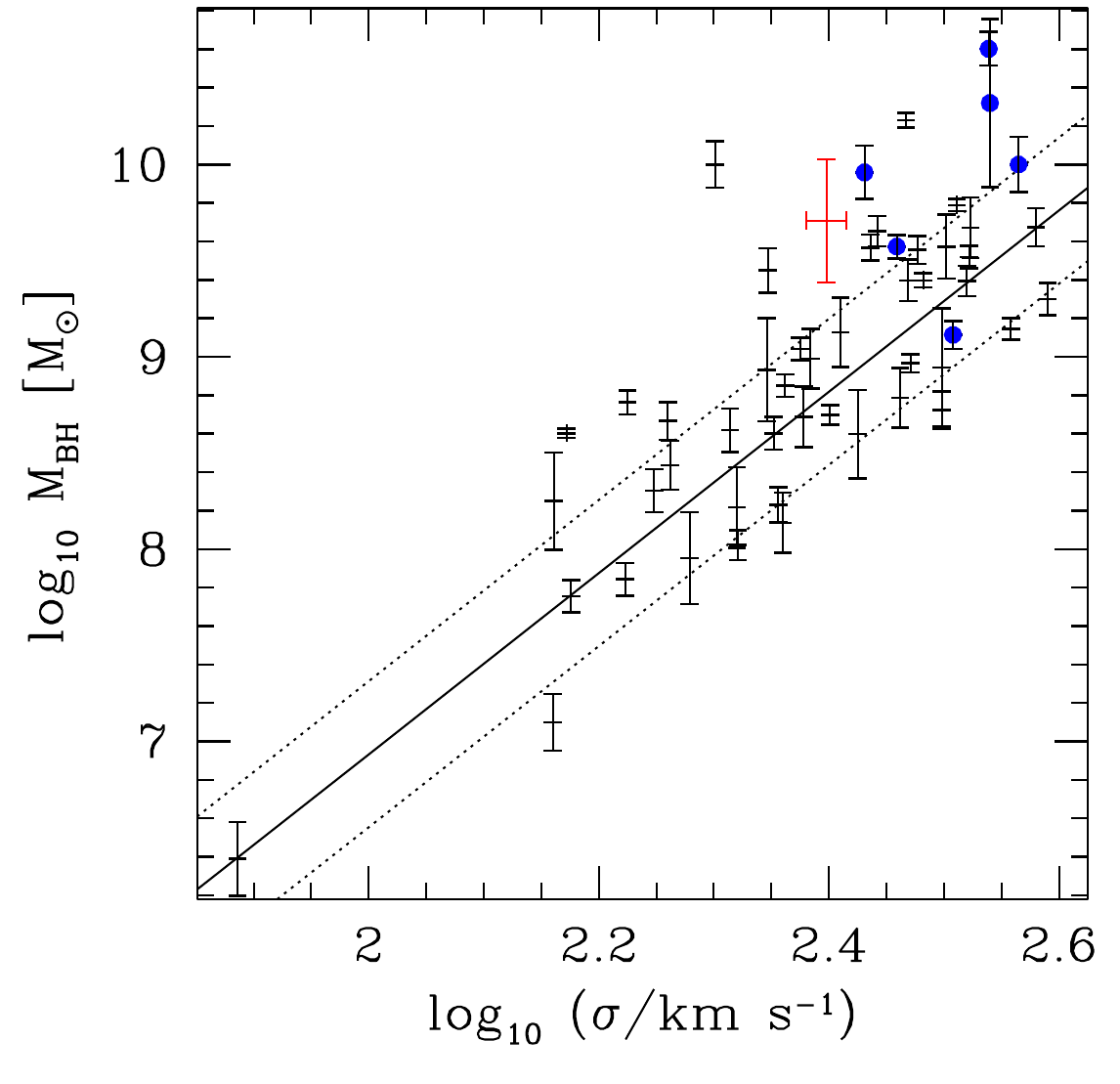} 
    \caption{Position of NGC 1272 (in red) on the $M_{\rm BH}$-$r_{\rm b}$ (left),
  on the $M_{\rm BH}$-$M_\ast$ (middle) and on the     
      $M_{\rm BH}$-$\sigma$ (right) relation. The black data points are from \citet{Rusli2013}, \citet{Saglia2016}, \citet{Thomas2016}, \citet{Mehrgan2019}, \citet{Neureiter2023b}, and \citet{deNicola2024}. The blue datapoints are core ellipticals with stellar mass larger than $10^{12} \solarmass$. While the galaxy follows the $M_{\rm BH}$-$r_{\rm b}$ given by \citet{Thomas2016} and the $M_{\rm BH}$-$M_\ast$ relation of \citet{Saglia2016} for the sample of CorePowerE, it deviates by a factor of 8.4 from the $M_{\rm BH}$-$\sigma$ relation of \citet{Saglia2016}, or by a factor of 1.8 of the $1\sigma$ error combined with the intrinsic scatter in the relation (shown by the dotted lines). }
    \label{Fig_MBHrbsigma}
\end{figure*}

We have presented a measurement of the size (0.45 \kiloparsec) of the
core of NGC 1272, based on the VIS image of the Perseus cluster taken
as part of the \Euclid ERO campaign. The dynamical modelling of the
stellar kinematics collected with the VIRUS spectrograph at the HET
allowed us to measure the mass $(5\pm 3)\expo{9} \solarmass$ of the BH
at the center of the galaxy.  While in line with expectations from the
$M_{\rm BH}$-$r_{\rm b}$ correlation of \citet{Thomas2016}, the
central surface brightness versus $M_{\rm BH}$ correlation of
\citet{Mehrgan2019}, and the $M_{\rm BH}$-$M_\ast$ relation of
\citet{Saglia2016}, the BH mass of NGC 1272 is a factor of 8 larger
than predicted by the $M_{\rm BH}$-$\sigma$ relation of
\citet{Saglia2016}, or 1.8 times the 1 sigma error combined with the
intrinsic scatter in the relation (see Fig.
\ref{Fig_MBHrbsigma}). This corroborates the conclusion that the
velocity dispersion is not the best indicator of the black hole mass
for core galaxies with stellar masses of the order of or larger than
$10^{12}\solarmass$: five out of the six galaxies with such a stellar
mass in Fig. \ref{Fig_MBHrbsigma} have BH masses larger than predicted
by the $M_{\rm BH}$-$\sigma$ relation. Therefore, the most efficient
and rapid method to search for galaxies harboring the most massive
black holes is to look for passive objects with large cores and low
central surface brightness. In the local Universe, a galaxy with a
core size of 1 \kiloparsec\ contains a black hole with a mass of
$10^{10} \solarmass$.

\begin{figure}[ht!]
    \centering
    \includegraphics[width=0.9\columnwidth]{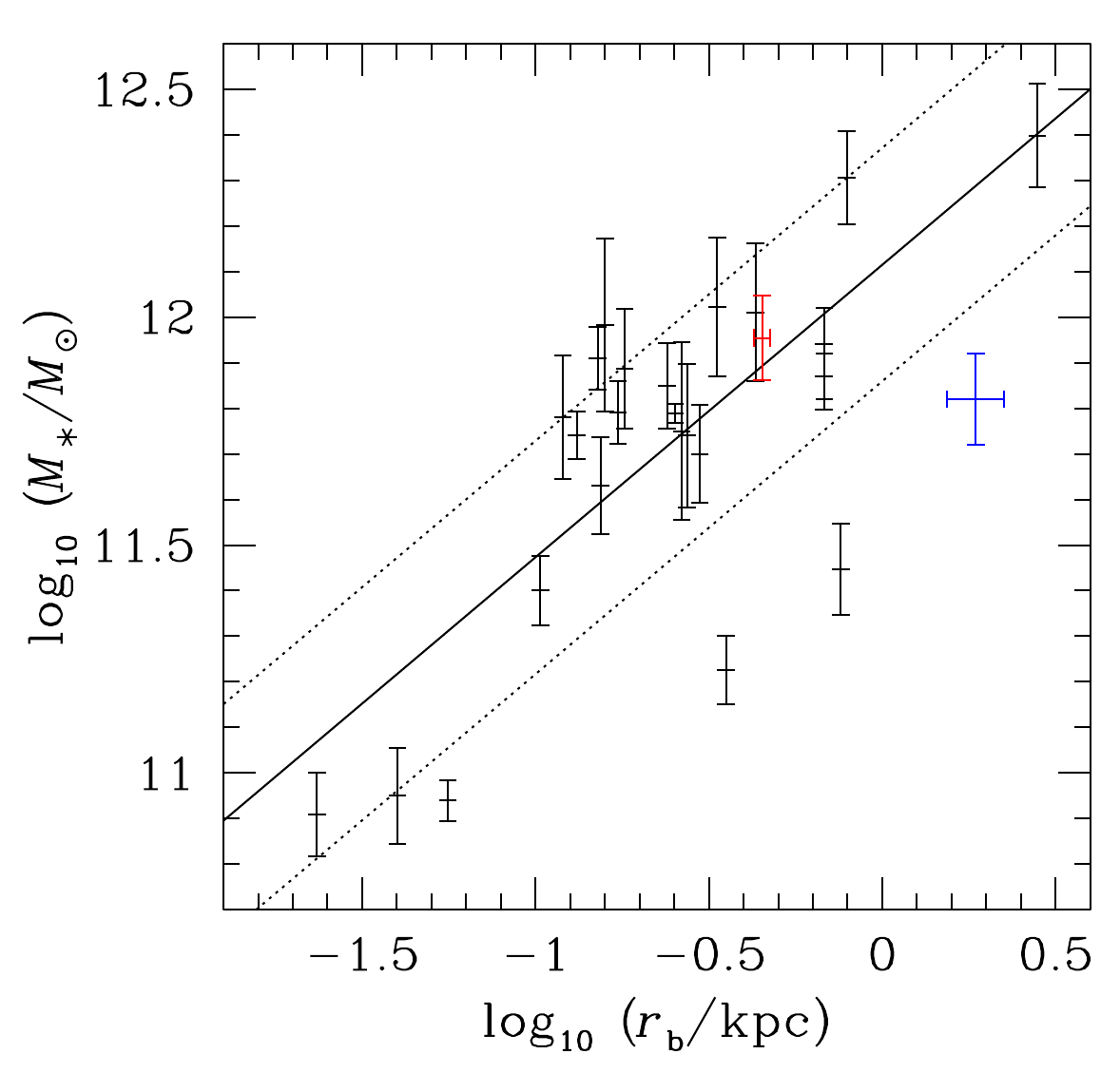} 
    \caption{Correlation between the core radius $r_{\rm b}$ and stellar mass
      $M_\ast$. The black datapoints are from \citet{Rusli2013}, \citet{Saglia2016},
      \citet{Thomas2016}, \citet{Mehrgan2019}, \citet{Neureiter2023b}, and \citet{deNicola2024}. NGC 1272 is shown in red. The blue cross shows the position of the BCG on the EDISCS cluster CL1216 \citep{Saglia2010}. The black solid line shows $\logten (M_\ast/M_\odot)=0.64 \logten (r_{\rm b}/\kiloparsec) +12.1$; the dotted lines show the $\pm 1 \sigma$ scatter (0.26 dex) in the relation.}
    \label{Fig_mstarrb}
\end{figure}

The Euclid VIS images in the \IE\ band deliver a PSF with FWHM
$\approx 0\arcsecf17$ with pixel size of $0\arcsecf1$, with a depth of
24.5 mag in the Wide Survey (at $10\sigma$ for extended sources) and 2
mag deeper in the Deep Survey. Near-infrared \YE, \JE, and \HE images
provide photometry with $0\arcsecf3$ pixels. Combined with
ground-based images, the surveys will deliver not only photometric
redshifts for each detected source, but also physical parameters, such as
stellar masses and sizes. At the end of the mission, the Wide Survey
will cover about 14,000 deg$^2$ of extragalactic sky, along with 50
deg$^2$ at the Deep Survey. This unprecedented dataset will allow us
to search for galaxies with cores larger than 2 \kiloparsec\ out to
redshift 1 (where they will subtend an angle of $0\arcsecf5$ on the
sky) as a function of stellar mass. We will establish up to which
redshift the correlation between $r_{\rm b}$ and total stellar mass exists
(see Fig. \ref{Fig_mstarrb}) and study its possible evolution with a
large statistical sample, indirectly probing the possible coevolution
of black holes and galaxy properties at the highest BH mass end. For
example, the blue cross in Fig. \ref{Fig_mstarrb} shows the position
of the BCG of the EDISCS cluster CL1216 at redshift 0.8
\citep{Saglia2010}. We measured the size of its core in the available
HST images, deriving 1.5 \kiloparsec\ (or $0\arcsecf5$) from a
core-S\'ersic fit, and 2.21 \kiloparsec\ (or $0\arcsecf7$) from a
Nuker fit. Such a core will be measurable in the VIS mosaics of the
Wide survey. With a stellar mass of $\logten M_\ast/M_\odot=11.82$ the
BCG appears to have a larger core than local core ellipticals of
similar mass. Using the local $r_{\rm b}$-$M_{\rm BH}$ relation, we
estimate that an HMBH with mass larger than $10^{10}\solarmass$ could
be already in place at such a high redshift in this
galaxy. Spectroscopic follow-up (possible at the Extremely Large
Telescope) of selected galaxies with similarly large and bright cores
will deliver the dynamical mass confirmation.

\begin{table}
  {\tiny
    \centering
    \caption{The parameters of the axisymmetric and triaxial dynamical modelling.}
    \begin{tabular}{lccccc}
    \hline
    \hline
\noalign{\smallskip}
    Model & $M_{\rm BH}$& $\Upsilon_{\ast,V}$ & $\logten \rho_{10}$ & $\gamma_{\rm DM}$ & $r_{\rm SOI}$\\
          &$[10^9 \solarmass]$ & $[\solarmass/\solarluminosity]$ & $[\solarmass\kiloparsec^{-3}]$ & & $[^{\prime\prime}]$\\
\noalign{\smallskip}
    \hline
    \noalign{\vskip 3pt}
    Axisymm.         &  5.1 $\pm$ 3.2 & 7.1 $\pm$ 1.5 & 7.2 $\pm$ 0.2   & 0$^{+0.3}$ & 1.24 $\pm$ 0.4\\
    Axisymm. Q3 & 1.8  & 8.9 & 7.3 & 0   & 0.8\\
    Axisymm. Q4 & 5.9  & 7.7 & 7.4 & 0   & 1.24\\ 
    Triaxial N      & 4.3  & 7.7 & 7.6 & 0.6 & 0.8\\
    Triaxial S      & 7.6  & 6.4 & 7.3 & 0.6 & 1.2\\   
    Prolate  N      & 7.6  & 6.4 & 7.1 & 0.0 & 1.5\\
    Prolate  S      & 1.0  & 5.1 & 7.0 & 0.2 & 0.7\\
    Spherical N     & 7.6  & 6.4 & 7.4 & 0.0 & 1.5\\
    Spherical S     & 10.9 & 3.9 & 7.8 & 0.8 & 1.7\\    
\noalign{\smallskip}
    \hline
    \end{tabular}
    \tablefoot{We list the model type (column 1), the black hole mass (column 2), the dynamically determined mass-to-light ratio (column 3), the logarithm of the dark matter density at 10 kpc (column 4), the inner slope of the dark matter density profile (column 5, bound to be larger or equal to 0), and the radius of the BH sphere of influence (column 6). }
    \label{Tab_DynMod}    
}
\end{table}

\begin{acknowledgements}
  \AckERO
  \AckEC
  RS, RB and MF acknowledge support by the Deutsches Zentrum f\"ur Luft-
  und Raumfahrt (DLR) grant 50 QE 1101. RS, RB, ML thank the Hobby Eberly
  Telescope (HET) project for allocating the observations and
  the technical support. The Hobby-Eberly Telescope is a joint project
  of the University of Texas at Austin, the Pennsylvania State
  University, Ludwig-Maximilians-Universität München, and Georg-August
  Universität Gottingen. The HET is named in honor of its principal
  benefactors, William P. Hobby and Robert E. Eberly. The HET
  Collaboration acknowledges the support and resources from the Texas
  Advanced Computing Center. We thank the Resident Astronomers and
  Telescope Operators at the HET for the skillful execution of our
  observations with VIRUS. We would like to acknowledge that the HET is
  built on Indigenous land. Moreover, we would like to acknowledge and
  pay our respects to the Carrizo \& Comecrudo, Coahuiltecan, Caddo,
  Tonkawa, Comanche, Lipan Apache, Alabama-Coushatta, Kickapoo, Tigua
  Pueblo, and all the American Indian and Indigenous Peoples and
  communities who have been or have become a part of these lands and
  territories in Texas, here on Turtle Island.\\
\end{acknowledgements}

\bibliographystyle{aa}

\begin{thebibliography}{55}
\expandafter\ifx\csname natexlab\endcsname\relax\def\natexlab#1{#1}\fi

\bibitem[{{Arakawa} {et~al.}(2019){Arakawa}, {Fabian}, \&
  {Walker}}]{Arakawa2019}
{Arakawa}, N., {Fabian}, A.~C., \& {Walker}, S.~A. 2019, \mnras, 488, 894

\bibitem[{{Bender} \& {Moellenhoff}(1987)}]{Bender1987}
{Bender}, R. \& {Moellenhoff}, C. 1987, \aap, 177, 71

\bibitem[{{Bender} {et~al.}(1989){Bender}, {Surma}, {Doebereiner},
  {Moellenhoff}, \& {Madejsky}}]{Bender1989}
{Bender}, R., {Surma}, P., {Doebereiner}, S., {Moellenhoff}, C., \& {Madejsky},
  R. 1989, \aap, 217, 35

\bibitem[{{Cappellari} \& {Copin}(2003)}]{Cappellari2003}
{Cappellari}, M. \& {Copin}, Y. 2003, \mnras, 342, 345

\bibitem[{{Choi} {et~al.}(2018){Choi}, {Somerville}, {Ostriker}, {Naab}, \&
  {Hirschmann}}]{Choi2018}
{Choi}, E., {Somerville}, R.~S., {Ostriker}, J.~P., {Naab}, T., \&
  {Hirschmann}, M. 2018, \apj, 866, 91

\bibitem[{{Cuillandre} {et~al.}(2024{\natexlab{a}}){Cuillandre}, {Bertin},
  {Bolzonella}, {Bouy}, {Gwyn}, {Isani}, {Kluge}, {Lai}, {Lan{\c{c}}on},
  {Lang}, {Laureijs}, {Saifollahi}, {Schirmer}, {Stone}, {Abdurro'uf},
  {Aghanim}, {Altieri}, {Annibali}, {Atek}, {Awad}, {Baes}, {Ba{\~n}ados},
  {Barrado}, {Belladitta}, {Belokurov}, {Boselli}, {Bournaud}, {Bovy},
  {Bowler}, {Buenadicha}, {Buitrago}, {Cantiello}, {Carollo}, {Codis},
  {Collins}, {Congedo}, {Dalessandro}, {de Lapparent}, {De Paolis}, {Diego},
  {Dimauro}, {Dinis}, {Dole}, {Duc}, {Erkal}, {Ezziati}, {Ferguson},
  {Ferr{\'e}-Mateu}, {Franco}, {Gavazzi}, {George}, {Gillard}, {Golden-Marx},
  {Goldman}, {Gonzalez}, {Habas}, {Hartley}, {Hatch}, {Kohley}, {Hoar},
  {Howell}, {Hunt}, {Jablonka}, {Jauzac}, {Kang}, {Knapen}, {Kneib}, {Kohley},
  {Kuzma}, {Larsen}, {Marchal}, {Mart{\'\i}n-Fleitas}, {Marcos-Arenal},
  {Marleau}, {Mart{\'\i}n}, {Massari}, {McConnachie}, {Meneghetti}, {Miluzio},
  {Miro Carretero}, {Miyatake}, {Mondelin}, {Montes}, {Mora}, {M{\"u}ller},
  {Nally}, {Noeske}, {Nucita}, {Oesch}, {Oguri}, {Peletier}, {Poulain},
  {Quilley}, {Racca}, {Rejkuba}, {Rhodes}, {Rocci}, {Rom{\'a}n}, {Sacquegna},
  {Saremi}, {Scaramella}, {Schinnerer}, {Serjeant}, {Sola}, {Sorce},
  {Tarsitano}, {Tereno}, {Toft}, {Tortora}, {Urbano}, {Venhola}, {Voggel},
  {Weaver}, {Xu}, {{\v{Z}}erjal}, {Z{\"o}ller}, {Andreon}, {Auricchio},
  {Baldi}, {Balestra}, {Bardelli}, {Basset}, {Bender}, {Bodendorf},
  {Branchini}, {Brau-Nogue}, {Brescia}, {Brinchmann}, {Camera}, {Capobianco},
  {Carbone}, {Carretero}, {Casas}, {Castander}, {Castellano}, {Cavuoti},
  {Cimatti}, {Conselice}, {Conversi}, {Copin}, {Courbin}, {Courtois},
  {Cropper}, {Cuby}, {Da Silva}, {Degaudenzi}, {Di Giorgio}, {Douspis},
  {Duncan}, {Dupac}, {Dusini}, {Fabricius}, {Farina}, {Farrens}, {Ferriol},
  {Fotopoulou}, {Frailis}, {Franceschi}, {Galeotta}, {Garilli}, {Gillis},
  {Giocoli}, {G{\'o}mez-Alvarez}, {Grazian}, {Grupp}, {Guzzo}, {Haugan},
  {Hoar}, {Hoekstra}, {Holmes}, {Hook}, {Hormuth}, {Hornstrup}, {Hudelot},
  {Jahnke}, {Jhabvala}, {Keih{\"a}nen}, {Kermiche}, {Kiessling}, {Kilbinger},
  {Kitching}, {Kubik}, {Kuijken}, {K{\"u}mmel}, {Kunz}, {Kurki-Suonio},
  {Lahav}, {Ligori}, {Lilje}, {Lindholm}, {Lloro}, {Maino}, {Maiorano},
  {Mansutti}, {Marggraf}, {Markovic}, {Martinet}, {Marulli}, {Massey},
  {Maurogordato}, {McCracken}, {Medinaceli}, {Mellier}, {Meylan}, {Mohr},
  {Moresco}, {Moscardini}, {Munari}, {Nakajima}, {Nichol}, {Niemi}, {Padilla},
  {Paltani}, {Pasian}, {Peacock}, {Pedersen}, {Percival}, {Pettorino}, {Pires},
  {Polenta}, {Poncet}, {Popa}, {Pozzetti}, {Raison}, {Rebolo}, {Refregier},
  {Renzi}, {Riccio}, {Rix}, {Romelli}, {Roncarelli}, {Rossetti}, {Saglia},
  {Sapone}, {Schneider}, {Schrabback}, {Secroun}, {Seidel}, {Serrano},
  {Sirignano}, {Sirri}, {Skottfelt}, {Stanco}, {Tallada-Cresp{\'\i}}, {Taylor},
  {Teplitz}, {Toledo-Moreo}, {Tsyganov}, {Tutusaus}, {Valentijn}, {Valenziano},
  {Vassallo}, {Verdoes Kleijn}, {Wang}, {Weller}, {Williams}, {Zamorani},
  {Zucca}, {Baccigalupi}, {Burigana}, {Casenove}, {Liebing}, {Scottez},
  {Simon}, \& {Scott}}]{Cuillandre2024}
{Cuillandre}, J.~C., {Bertin}, E., {Bolzonella}, M., {et~al.}
  2024{\natexlab{a}}, \aap, submitted, arXiv:2405.13496

\bibitem[{{Cuillandre} {et~al.}(2024{\natexlab{b}}){Cuillandre}, {Bolzonella},
  {Boselli}, {Marleau}, {Mondelin}, {Sorce}, {Stone}, {Buitrago}, {Cantiello},
  {George}, {Hatch}, {Quilley}, {Mannucci}, {Saifollahi},
  {S{\'a}nchez-Janssen}, {Tarsitano}, {Tortora}, {Xu}, {Bouy}, {Gwyn}, {Kluge},
  {Lan{\c{c}}on}, {Laureijs}, {Schirmer}, {Abdurro'uf}, {Awad}, {Baes},
  {Bournaud}, {Carollo}, {Codis}, {Conselice}, {De Lapparent}, {Duc},
  {Ferr{\'e}-Mateu}, {Gillard}, {Golden-Marx}, {Jablonka}, {Habas}, {Hunt},
  {Mei}, {Miville-Desch{\^e}nes}, {Montes}, {Nersesian}, {Peletier}, {Poulain},
  {Scaramella}, {Scialpi}, {Sola}, {Stephan}, {Ulivi}, {Urbano}, {Z{\"o}ller},
  {Aghanim}, {Altieri}, {Amara}, {Andreon}, {Auricchio}, {Baldi}, {Balestra},
  {Bardelli}, {Bender}, {Bodendorf}, {Bonino}, {Branchini}, {Brescia},
  {Brinchmann}, {Camera}, {Capobianco}, {Carbone}, {Carretero}, {Casas},
  {Castander}, {Castellano}, {Cavuoti}, {Cimatti}, {Congedo}, {Conversi},
  {Copin}, {Courbin}, {Courtois}, {Cropper}, {Da Silva}, {Degaudenzi}, {Di
  Giorgio}, {Dinis}, {Douspis}, {Dubath}, {Duncan}, {Dupac}, {Dusini},
  {Farina}, {Farrens}, {Ferriol}, {Fotopoulou}, {Frailis}, {Franceschi},
  {Galeotta}, {Gillis}, {Giocoli}, {G{\'o}mez-Alvarez}, {Grazian}, {Grupp},
  {Guzzo}, {Haugan}, {Hoar}, {Hoekstra}, {Holmes}, {Hook}, {Hormuth},
  {Hornstrup}, {Hudelot}, {Jahnke}, {Jhabvala}, {Keih{\"a}nen}, {Kermiche},
  {Kiessling}, {Kilbinger}, {Kitching}, {Kohley}, {Kubik}, {Kuijken},
  {K{\"u}mmel}, {Kunz}, {Kurki-Suonio}, {Lahav}, {Le Mignant}, {Ligori},
  {Lilje}, {Lindholm}, {Lloro}, {Maino}, {Maiorano}, {Mansutti}, {Marggraf},
  {Markovic}, {Martinet}, {Marulli}, {Massey}, {Maurogordato}, {McCracken},
  {Medinaceli}, {Melchior}, {Mellier}, {Meneghetti}, {Merlin}, {Meylan},
  {Mohr}, {Moresco}, {Moscardini}, {Nakajima}, {Nichol}, {Niemi}, {Padilla},
  {Paltani}, {Pasian}, {Pedersen}, {Percival}, {Pettorino}, {Pires}, {Polenta},
  {Poncet}, {Popa}, {Pozzetti}, {Raison}, {Renzi}, {Rhodes}, {Riccio},
  {Romelli}, {Roncarelli}, {Saglia}, {Sapone}, {Schneider}, {Schrabback},
  {Secroun}, {Seidel}, {Serrano}, {Sirignano}, {Sirri}, {Skottfelt}, {Stanco},
  {Tallada-Cresp{\'\i}}, {Taylor}, {Teplitz}, {Tereno}, {Toledo-Moreo},
  {Tutusaus}, {Valentijn}, {Valenziano}, {Vassallo}, {Verdoes Kleijn}, {Wang},
  {Weller}, {Zucca}, {Biviano}, {Burigana}, {Castignani}, {De Lucia},
  {Scottez}, {Mora}, {Simon}, {Mart{\'\i}n-Fleitas}, \& {Scott}}]{Perseus2024}
{Cuillandre}, J.~C., {Bolzonella}, M., {Boselli}, A., {et~al.}
  2024{\natexlab{b}}, \aap, submitted, arXiv:2405.13501

\bibitem[{{de Nicola} {et~al.}(2022{\natexlab{a}}){de Nicola}, {Neureiter},
  {Thomas}, {Saglia}, \& {Bender}}]{deNicola2022b}
{de Nicola}, S., {Neureiter}, B., {Thomas}, J., {Saglia}, R.~P., \& {Bender},
  R. 2022{\natexlab{a}}, \mnras, 517, 3445

\bibitem[{{de Nicola} {et~al.}(2020){de Nicola}, {Saglia}, {Thomas}, {Dehnen},
  \& {Bender}}]{deNicola2020}
{de Nicola}, S., {Saglia}, R.~P., {Thomas}, J., {Dehnen}, W., \& {Bender}, R.
  2020, \mnras, 496, 3076

\bibitem[{{de Nicola} {et~al.}(2022{\natexlab{b}}){de Nicola}, {Saglia},
  {Thomas}, {Pulsoni}, {Kluge}, {Bender}, {Valenzuela}, \&
  {Remus}}]{deNicola2022a}
{de Nicola}, S., {Saglia}, R.~P., {Thomas}, J., {et~al.} 2022{\natexlab{b}},
  \apj, 933, 215

\bibitem[{{de Nicola} {et~al.}(2024){de Nicola}, {Thomas}, {Saglia}, {Snigula},
  {Kluge}, \& {Bender}}]{deNicola2024}
{de Nicola}, S., {Thomas}, J., {Saglia}, R.~P., {et~al.} 2024, \mnras, 530,
  1035

\bibitem[{{de Rijcke} {et~al.}(2009){de Rijcke}, {Penny}, {Conselice},
  {Valcke}, \& {Held}}]{deRijke2009}
{de Rijcke}, S., {Penny}, S.~J., {Conselice}, C.~J., {Valcke}, S., \& {Held},
  E.~V. 2009, \mnras, 393, 798

\bibitem[{{de Vaucouleurs} {et~al.}(1991){de Vaucouleurs}, {de Vaucouleurs},
  {Corwin}, {Buta}, {Paturel}, \& {Fouque}}]{RC3}
{de Vaucouleurs}, G., {de Vaucouleurs}, A., {Corwin}, Herold~G., J., {et~al.}
  1991, {Third Reference Catalogue of Bright Galaxies} (Springer)

\bibitem[{{Erwin}(2015)}]{Erwin2015}
{Erwin}, P. 2015, \apj, 799, 226

\bibitem[{{Euclid Collaboration: Cropper} {et~al.}(2024){Euclid Collaboration:
  Cropper}, {Al-Bahlawan}, {Amiaux}, {Awan}, {Azzollini}, {Benson}, {Berthe},
  {Boucher}, {Bozzo}, {Brockley-Blatt}, {Candini}, {Cara}, {Chaudery}, {Cole},
  {Danto}, {Denniston}, {Di Giorgio}, {Dryer}, {Endicott}, {Dubois}, {Farina},
  {Galli}, {Genolet}, {Gow}, {Guttridge}, {Hailey}, {Hall}, {Harper},
  {Holland}, {Horeau}, {Hu}, {King}, {James}, {Larcheveque}, {Khalil},
  {Lawrenson}, {Liebing}, {Martignac}, {McCracken}, {Murray}, {Nakajima},
  {Niemi}, {Pendem}, {Paltani}, {Philippon}, {Pool}, {Plana}, {Pottinger},
  {Racca}, {Rousseau}, {Ruane}, {Salatti}, {Salvignol}, {Sciortino}, {Short},
  {Liu}, {Skottfelt}, {Swindells}, {Smit}, {Szafraniec}, {Thomas}, {Thomas},
  {Tommasi}, {Winter}, {Tosti}, {Visticot}, {Walton}, {Willis}, {Mora},
  {Kohley}, {Massey}, {Nightingale}, {Kitching}, {Hoekstra}, {Aghanim},
  {Altieri}, {Amara}, {Andreon}, {Auricchio}, {Aussel}, {Baldi}, {Balestra},
  {Bardelli}, {Basset}, {Bender}, {Bodendorf}, {Boenke}, {Bonino}, {Branchini},
  {Brescia}, {Brinchmann}, {Camera}, {Capobianco}, {Carbone}, {Cardone},
  {Carretero}, {Casas}, {Casas}, {Castander}, {Castellano}, {Cavuoti},
  {Cimatti}, {Congedo}, {Conselice}, {Conversi}, {Copin}, {Courbin},
  {Courtois}, {Cuby}, {Cuillandre}, {Da Silva}, {Degaudenzi}, {Dinis},
  {Dolding}, {Douspis}, {Duncan}, {Dupac}, {Dusini}, {Ealet}, {Fabricius},
  {Farrens}, {Ferriol}, {Fosalba}, {Fotopoulou}, {Frailis}, {Franceschi},
  {Franzetti}, {Frugier}, {Fumana}, {Galeotta}, {Garilli}, {Gillard}, {Gillis},
  {Giocoli}, {G{\'o}mez-Alvarez}, {Granett}, {Grazian}, {Grupp}, {Guzzo},
  {Haugan}, {Herent}, {Hoar}, {Holliman}, {Hook}, {Hormuth}, {Hornstrup},
  {Hudelot}, {Jahnke}, {Jhabvala}, {Joachimi}, {Keih{\"a}nen}, {Kermiche},
  {Kilbinger}, {Kubik}, {Kuijken}, {K{\"u}mmel}, {Kunz}, {Kurki-Suonio},
  {Lahav}, {Laureijs}, {Ligori}, {Lilje}, {Lindholm}, {Lloro}, {Alvarez},
  {Maino}, {Maiorano}, {Mansutti}, {Marggraf}, {Martinet}, {Marulli},
  {Masters}, {Maurogordato}, {Medinaceli}, {Mei}, {Melchior}, {Mellier},
  {Meneghetti}, {Merlin}, {Meylan}, {Miller}, {Mohr}, {Moresco}, {Moscardini},
  {Nichol}, {Nutma}, {Padilla}, {Paech}, {Pasian}, {Peacock}, {Pedersen},
  {Percival}, {Pettorino}, {Pires}, {Polenta}, {Poncet}, {Popa}, {Pozzetti},
  {Raison}, {Rebolo}, {Refregier}, {Renzi}, {Riccio}, {Rix}, {Romelli},
  {Roncarelli}, {Rosset}, {Rossetti}, {Rottgering}, {Saglia}, {Sapone},
  {Sauvage}, {Scaramella}, {Schirmer}, {Schneider}, {Schrabback}, {Secroun},
  {Seidel}, {Serrano}, {Sirignano}, {Sirri}, {Stanco}, {Starck},
  {Tallada-Cresp{\'\i}}, {Tavagnacco}, {Taylor}, {Teplitz}, {Tereno},
  {Toledo-Moreo}, {Torradeflot}, {Tutusaus}, {Valentijn}, {Valenziano},
  {Vassallo}, {Verdoes Kleijn}, {Veropalumbo}, {Wachter}, {Wang}, {Weller},
  {Zamorani}, {Zoubian}, {Zucca}, {Baccigalupi}, {Bernardeau}, {Biviano},
  {Bolzonella}, {Boucaud}, {Burigana}, {Calabrese}, {Casenove},
  {Colodro-Conde}, {Crocce}, {De Lucia}, {Di Ferdinando}, {Escartin Vigo},
  {Fabbian}, {Farinelli}, {Finelli}, {George}, {Gracia-Carpio}, {Ili{\'c}},
  {Israel}, {Mainetti}, {Marcin}, {Martinelli}, {Mauri}, {Neissner},
  {Nguyen-Kim}, {Pezzotta}, {P{\"o}ntinen}, {Porciani}, {Sakr}, {Scottez},
  {Sefusatti}, {Tenti}, {Viel}, {Wiesmann}, {Akrami}, {Allevato}, {Anselmi},
  {Aubourg}, {Ballardini}, {Bertacca}, {Bethermin}, {Blanchard}, {Blot},
  {Borgani}, {Borlaff}, {Bruton}, {Cabanac}, {Calabro}, {Calderone},
  {Canas-Herrera}, {Cappi}, {Carvalho}, {Castignani}, {Castro}, {Chambers},
  {Chary}, {Contarini}, {Cooray}, {Cordes}, {Costanzi}, {Cucciati}, {Davini},
  {De Caro}, {Desprez}, {D{\'\i}az-S{\'a}nchez}, {Di Domizio}, {Dole},
  {Escoffier}, {Ferrari}, {Ferreira}, {Ferrero}, {Finoguenov}, {Fontana},
  {Fornari}, {Gabarra}, {Ganga}, {Garc{\'\i}a-Bellido}, {Gautard}, {Gaztanaga},
  {Giacomini}, {Gianotti}, {Gozaliasl}, {Gregorio}, {Hall}, {Hartley},
  {Hildebrandt}, {Hjorth}, {Huertas-Company}, {Ilbert}, {Joudaki}, {Kajava},
  {Kansal}, {Karagiannis}, {Kirkpatrick}, {Lacasa}, {Le Graet}, {Legrand},
  {Libet}, {Loureiro}, {Macias-Perez}, {Magliocchetti}, {Mancini}, {Mannucci},
  {Maoli}, {Martins}, {Matthew}, {Maurin}, {McPartland}, {Metcalf},
  {Migliaccio}, {Miluzio}, {Monaco}, {Moretti}, {Morgante}, {Nadathur},
  {Walton}, {Odier}, {Oguri}, {Patrizii}, {Popa}, {Potter}, {Pourtsidou},
  {Reimberg}, {Risso}, {Rocci}, {Rollins}, {Rusholme}, {Sahl{\'e}n},
  {S{\'a}nchez}, {Scarlata}, {Schaye}, {Schewtschenko}, {Schneider},
  {Schultheis}, {Sereno}, {Shankar}, {Sikkema}, {Silvestri}, {Simon}, {Spurio
  Mancini}, {Stadel}, {Stanford}, {Steinwagner}, {Tanidis}, {Tao}, {Tessore},
  {Testera}, {Tewes}, {Teyssier}, {Toft}, {Tosi}, {Troja}, {Tucci}, {Valieri},
  {Valiviita}, {Vergani}, {Vernizzi}, {Verza}, {Vielzeuf}, {Weaver}, {Zalesky},
  {Zinchenko}, {Archidiacono}, {Atrio-Barandela}, {Bouvard}, {Caro}, {Dimauro},
  {Duc}, {Fang}, {Ferguson}, {Gasparetto}, {Gutierrez},
  {Kova\{{\v{c}}\}i{\'c}}, {Kruk}, {Le Brun}, {Liaudat}, {Montoro}, {Murray},
  {Pagano}, {Paoletti}, {Sarpa}, {Viitanen}, {Lesgourgues}, \&
  {Mart{\'\i}n-Fleitas}}]{Cropper2024}
{Euclid Collaboration: Cropper}, M., {Al-Bahlawan}, A., {Amiaux}, J., {et~al.}
  2024, \aap, submitted, arXiv:2405.13492

\bibitem[{{Euclid Collaboration: Jahnke} {et~al.}(2024){Euclid Collaboration:
  Jahnke}, {Gillard}, {Schirmer}, {Ealet}, {Maciaszek}, {Prieto}, {Barbier},
  {Bonoli}, {Corcione}, {Dusini}, {Grupp}, {Hormuth}, {Ligori}, {Martin},
  {Morgante}, {Padilla}, {Toledo-Moreo}, {Trifoglio}, {Valenziano}, {Bender},
  {Castander}, {Garilli}, {Lilje}, {Rix}, {Auricchio}, {Balestra}, {Barriere},
  {Battaglia}, {Berthe}, {Bodendorf}, {Boenke}, {Bon}, {Bonnefoi}, {Caillat},
  {Capobianco}, {Carle}, {Casas}, {Cho}, {Costille}, {Ducret}, {Ferriol},
  {Franceschi}, {Gimenez}, {Holmes}, {Hornstrup}, {Jhabvala}, {Kohley},
  {Kubik}, {Laureijs}, {Le Mignant}, {Lloro}, {Medinaceli}, {Mellier},
  {Polenta}, {Racca}, {Renzi}, {Salvignol}, {Secroun}, {Seidel}, {Seiffert},
  {Sirignano}, {Sirri}, {Strada}, {Smadja}, {Stanco}, {Wachter}, {Anselmi},
  {Borsato}, {Caillat}, {Cogato}, {Colodro-Conde}, {Crouzet}, {Conforti},
  {D'Alessandro}, {Copin}, {Cuillandre}, {Davies}, {Davini}, {Derosa}, {Diaz},
  {Di Domizio}, {Di Ferdinando}, {Farinelli}, {Ferrari}, {Fornari}, {Gabarra},
  {Gutierrez}, {Giacomini}, {Lagier}, {Gianotti}, {Krause}, {Madrid},
  {Laudisio}, {Macias-Perez}, {Naletto}, {Niclas}, {Marpaud}, {Mauri}, {da
  Silva}, {Passalacqua}, {Paterson}, {Patrizii}, {Risso}, {Solheim},
  {Scodeggio}, {Stassi}, {Steinwagner}, {Tenti}, {Testera}, {Travaglini},
  {Tosi}, {Troja}, {Tubio}, {Valieri}, {Vescovi}, {Ventura}, {Aghanim},
  {Altieri}, {Amara}, {Amiaux}, {Andreon}, {Aussel}, {Baldi}, {Bardelli},
  {Basset}, {Bonchi}, {Bonino}, {Branchini}, {Brescia}, {Brinchmann}, {Camera},
  {Carbone}, {Cardone}, {Carretero}, {Casas}, {Castellano}, {Cavuoti},
  {Chabaud}, {Cimatti}, {Congedo}, {Conselice}, {Conversi}, {Courbin},
  {Courtois}, {Cropper}, {Cuby}, {Da Silva}, {Degaudenzi}, {Di Giorgio},
  {Dinis}, {Douspis}, {Dubath}, {Duncan}, {Dupac}, {Fabricius}, {Farina},
  {Farrens}, {Faustini}, {Fosalba}, {Fotopoulou}, {Fourmanoit}, {Frailis},
  {Franzetti}, {Galeotta}, {Gillis}, {Giocoli}, {G{\'o}mez-Alvarez}, {Granett},
  {Grazian}, {Guzzo}, {Hailey}, {Haugan}, {Hoar}, {Hoekstra}, {Hook},
  {Hudelot}, {Joachimi}, {Keih{\"a}nen}, {Kermiche}, {Kiessling}, {Kilbinger},
  {Kitching}, {K{\"u}mmel}, {Kunz}, {Kurki-Suonio}, {Lahav}, {Lindholm},
  {Alvarez}, {Maino}, {Maiorano}, {Mansutti}, {Marggraf}, {Markovic},
  {Martignac}, {Martinet}, {Marulli}, {Massey}, {Masters}, {Maurogordato},
  {McCracken}, {Mei}, {Melchior}, {Meneghetti}, {Merlin}, {Meylan}, {Mohr},
  {Moresco}, {Moscardini}, {Nakajima}, {Nichol}, {Niemi}, {Nutma}, {Paech},
  {Paltani}, {Pasian}, {Peacock}, {Pedersen}, {Percival}, {Pettorino}, {Pires},
  {Poncet}, {Popa}, {Pozzetti}, {Raison}, {Rebolo}, {Refregier}, {Rhodes},
  {Riccio}, {Romelli}, {Roncarelli}, {Rosset}, {Rossetti}, {Rottgering},
  {Saglia}, {Sapone}, {Sauvage}, {Scaramella}, {Schneider}, {Schrabback},
  {Serrano}, {Tallada-Cresp{\'\i}}, {Tavagnacco}, {Taylor}, {Teplitz},
  {Tereno}, {Torradeflot}, {Tutusaus}, {Vassallo}, {Verdoes Kleijn},
  {Veropalumbo}, {Vibert}, {Wang}, {Weller}, {Zacchei}, {Zamorani}, {Zerbi},
  {Zoubian}, {Zucca}, {Appleton}, {Baccigalupi}, {Biviano}, {Bolzonella},
  {Boucaud}, {Bozzo}, {Burigana}, {Calabrese}, {Casenove}, {Crocce}, {De
  Lucia}, {Escartin Vigo}, {Fabbian}, {Finelli}, {George}, {Gracia-Carpio},
  {Ili{\'c}}, {Liebing}, {Liu}, {Mainetti}, {Marcin}, {Martinelli}, {Morris},
  {Neissner}, {Pezzotta}, {P{\"o}ntinen}, {Porciani}, {Sakr}, {Scottez},
  {Sefusatti}, {Viel}, {Wiesmann}, {Akrami}, {Allevato}, {Aubourg},
  {Ballardini}, {Bertacca}, {Bethermin}, {Blanchard}, {Blot}, {Borgani},
  {Borlaff}, {Bruton}, {Cabanac}, {Calabro}, {Calderone}, {Canas-Herrera},
  {Cappi}, {Carvalho}, {Castignani}, {Castro}, {Chambers}, {Charles}, {Chary},
  {Colbert}, {Contarini}, {Contini}, {Cooray}, {Costanzi}, {Cucciati}, {De
  Caro}, {de la Torre}, {Desprez}, {D{\'\i}az-S{\'a}nchez}, {Dole},
  {Escoffier}, {Ferreira}, {Ferrero}, {Finoguenov}, {Fontana}, {Ganga},
  {Garc{\'\i}a-Bellido}, {Gautard}, {Gaztanaga}, {Gozaliasl}, {Gregorio},
  {Hall}, {Hartley}, {Hemmati}, {Hildebrandt}, {Hjorth}, {Hosseini},
  {Huertas-Company}, {Ilbert}, {Jacobson}, {Joudaki}, {Kajava}, {Kansal},
  {Karagiannis}, {Kirkpatrick}, {Lacasa}, {Le Brun}, {Le Graet}, {Legrand},
  {Libet}, {Liu}, {Loureiro}, {Magliocchetti}, {Mancini}, {Mannucci}, {Maoli},
  {Martins}, {Matthew}, {Maurin}, {McPartland}, {Metcalf}, {Migliaccio},
  {Miluzio}, {Monaco}, {Moretti}, {Nadathur}, {Nicastro}, {Walton}, {Odier},
  {Oguri}, {Popa}, {Potter}, {Pourtsidou}, {Rocci}, {Rollins}, {Rusholme},
  {Sahl{\'e}n}, {S{\'a}nchez}, {Scarlata}, {Schaye}, {Schewtschenko},
  {Schneider}, {Schultheis}, {Sereno}, {Shankar}, {Shulevski}, {Sikkema},
  {Silvestri}, {Simon}, {Spurio Mancini}, {Stadel}, {Stanford}, {Tanidis},
  {Tao}, {Tessore}, {Teyssier}, {Toft}, {Tucci}, {Valiviita}, {Vergani},
  {Vernizzi}, {Verza}, {Vielzeuf}, {Weaver}, {Zalesky}, {Zinchenko},
  {Archidiacono}, {Atrio-Barandela}, {Bennett}, {Bouvard}, {Caro}, {Conseil},
  {Dimauro}, {Duc}, {Fang}, {Ferguson}, {Gasparetto}, {Kova\{{\v{c}}\}i{\'c}},
  {Kruk}, {Le Brun}, {Liaudat}, {Montoro}, {Mora}, {Murray}, {Pagano},
  {Paoletti}, {Radovich}, {Sarpa}, {Tommasi}, {Viitanen}, {Lesgourgues},
  {Levi}, \& {Mart{\'\i}n-Fleitas}}]{Janke2024}
{Euclid Collaboration: Jahnke}, K., {Gillard}, W., {Schirmer}, M., {et~al.}
  2024, \aap, submitted, arXiv:2405.13493

\bibitem[{{Euclid Collaboration: Mellier} {et~al.}(2024){Euclid Collaboration:
  Mellier}, {Abdurro'uf}, {Acevedo Barroso}, {Ach{\'u}carro}, {Adamek}, {Adam},
  {Addison}, {Aghanim}, {Aguena}, {Ajani}, {Akrami}, {Al-Bahlawan}, {Alavi},
  {Albuquerque}, {Alestas}, {Alguero}, {Allaoui}, {Allen}, {Allevato},
  {Alonso-Tetilla}, {Altieri}, {Alvarez-Candal}, {Amara}, {Amendola}, {Amiaux},
  {Andika}, {Andreon}, {Andrews}, {Angora}, {Angulo}, {Annibali}, {Anselmi},
  {Anselmi}, {Arcari}, {Archidiacono}, {Aric{\`o}}, {Arnaud}, {Arnouts},
  {Asgari}, {Asorey}, {Atayde}, {Atek}, {Atrio-Barandela}, {Aubert}, {Aubourg},
  {Auphan}, {Auricchio}, {Aussel}, {Aussel}, {Avelino}, {Avgoustidis}, {Avila},
  {Awan}, {Azzollini}, {Baccigalupi}, {Bachelet}, {Bacon}, {Baes}, {Bagley},
  {Bahr-Kalus}, {Balaguera-Antolinez}, {Balbinot}, {Balcells}, {Baldi},
  {Baldry}, {Balestra}, {Ballardini}, {Ballester}, {Balogh}, {Ba{\~n}ados},
  {Barbier}, {Bardelli}, {Barreiro}, {Barriere}, {Barros}, {Barthelemy},
  {Bartolo}, {Basset}, {Battaglia}, {Battisti}, {Baugh}, {Baumont},
  {Bazzanini}, {Beaulieu}, {Beckmann}, {Belikov}, {Bel}, {Bellagamba}, {Bella},
  {Bellini}, {Benabed}, {Bender}, {Benevento}, {Bennett}, {Benson},
  {Bergamini}, {Bermejo-Climent}, {Bernardeau}, {Bertacca}, {Berthe},
  {Berthier}, {Bethermin}, {Beutler}, {Bevillon}, {Bhargava}, {Bhatawdekar},
  {Bisigello}, {Biviano}, {Blake}, {Blanchard}, {Blazek}, {Blot}, {Bosco},
  {Bodendorf}, {Boenke}, {B{\"o}hringer}, {Bolzonella}, {Bonchi}, {Bonici},
  {Bonino}, {Bonino}, {Bonvin}, {Bon}, {Booth}, {Borgani}, {Borlaff},
  {Borsato}, {Bosco}, {Bose}, {Botticella}, {Boucaud}, {Bouche}, {Boucher},
  {Boutigny}, {Bouvard}, {Bouy}, {Bowler}, {Bozza}, {Bozzo}, {Branchini},
  {Brau-Nogue}, {Brekke}, {Bremer}, {Brescia}, {Breton}, {Brinchmann},
  {Brinckmann}, {Brockley-Blatt}, {Brodwin}, {Brouard}, {Brown}, {Bruton},
  {Bucko}, {Buddelmeijer}, {Buenadicha}, {Buitrago}, {Burger}, {Burigana},
  {Busillo}, {Busonero}, {Cabanac}, {Cabayol-Garcia}, {Cagliari}, {Caillat},
  {Caillat}, {Calabrese}, {Calabro}, {Calderone}, {Calura}, {Camacho Quevedo},
  {Camera}, {Campos}, {Canas-Herrera}, {Candini}, {Cantiello}, {Capobianco},
  {Cappellaro}, {Cappelluti}, {Cappi}, {Caputi}, {Cara}, {Carbone}, {Cardone},
  {Carella}, {Carlberg}, {Carle}, {Carminati}, {Caro}, {Carrasco}, {Carretero},
  {Carrilho}, {Carron Duque}, {Carry}, {Carvalho}, {Carvalho}, {Casas},
  {Casas}, {Casenove}, {Casey}, {Cassata}, {Castander}, {Castelao},
  {Castellano}, {Castiblanco}, {Castignani}, {Castro}, {Cavet}, {Cavuoti},
  {Chabaud}, {Chambers}, {Charles}, {Charlot}, {Chartab}, {Chary}, {Chaumeil},
  {Cho}, {Chon}, {Ciancetta}, {Ciliegi}, {Cimatti}, {Cimino}, {Cioni},
  {Claydon}, {Cleland}, {Cl{\'e}ment}, {Clements}, {Clerc}, {Clesse}, {Codis},
  {Cogato}, {Colbert}, {Cole}, {Coles}, {Collett}, {Collins}, {Colodro-Conde},
  {Colombo}, {Combes}, {Conforti}, {Congedo}, {Conseil}, {Conselice},
  {Contarini}, {Contini}, {Conversi}, {Cooray}, {Copin}, {Corasaniti},
  {Corcho-Caballero}, {Corcione}, {Cordes}, {Corpace}, {Correnti}, {Costanzi},
  {Costille}, {Courbin}, {Courcoult Mifsud}, {Courtois}, {Cousinou}, {Covone},
  {Cowell}, {Cragg}, {Cresci}, {Cristiani}, {Crocce}, {Cropper}, {E Crouzet},
  {Csizi}, {Cuby}, {Cucchetti}, {Cucciati}, {Cuillandre}, {Cunha}, {Cuozzo},
  {Daddi}, {D'Addona}, {Dafonte}, {Dagoneau}, {Dalessandro}, {Dalton},
  {D'Amico}, {Dannerbauer}, {Danto}, {Das}, {Da Silva}, {da Silva}, {Daste},
  {Davies}, {Davini}, {de Boer}, {Decarli}, {De Caro}, {Degaudenzi}, {Degni},
  {de Jong}, {de la Bella}, {de la Torre}, {Delhaise}, {Delley}, {Delucchi},
  {De Lucia}, {Denniston}, {De Paolis}, {De Petris}, {Derosa}, {Desai},
  {Desjacques}, {Despali}, {Desprez}, {De Vicente-Albendea}, {Deville}, {Dias},
  {D{\'\i}az-S{\'a}nchez}, {Diaz}, {Di Domizio}, {Diego}, {Di Ferdinando}, {Di
  Giorgio}, {Dimauro}, {Dinis}, {Dolag}, {Dolding}, {Dole}, {Dom{\'\i}nguez
  S{\'a}nchez}, {Dor{\'e}}, {Dournac}, {Douspis}, {Dreihahn}, {Droge}, {Dryer},
  {Dubath}, {Duc}, {Ducret}, {Duffy}, {Dufresne}, {Duncan}, {Dupac}, {Duret},
  {Durrer}, {Durret}, {Dusini}, {Ealet}, {Eggemeier}, {Eisenhardt}, {Elbaz},
  {Elkhashab}, {Ellien}, {Endicott}, {Enia}, {Erben}, {Escartin Vigo},
  {Escoffier}, {Escudero Sanz}, {Essert}, {Ettori}, {Ezziati}, {Fabbian},
  {Fabricius}, {Fang}, {Farina}, {Farina}, {Farinelli}, {Farrens}, {Faustini},
  {Feltre}, {Ferguson}, {Ferrando}, {Ferrari}, {Ferr{\'e}-Mateu}, {Ferreira},
  {Ferreras}, {Ferrero}, {Ferriol}, {Ferruit}, {Filleul}, {Finelli},
  {Finkelstein}, {Finoguenov}, {Fiorini}, {Flentge}, {Focardi}, {Fonseca},
  {Fontana}, {Fontanot}, {Fornari}, {Fosalba}, {Fossati}, {Fotopoulou},
  {Fouchez}, {Fourmanoit}, {Frailis}, {Fraix-Burnet}, {Franceschi}, {Franco},
  {Franzetti}, {Freihoefer}, {Frittoli}, {Frugier}, {Frusciante}, {Fumagalli},
  {Fumagalli}, {Fumana}, {Fu}, {Gabarra}, {Galeotta}, {Galluccio}, {Ganga},
  {Gao}, {Garc{\'\i}a-Bellido}, {Garcia}, {Gardner}, {Garilli},
  {Gaspar-Venancio}, {Gasparetto}, {Gautard}, {Gavazzi}, {Gaztanaga},
  {Genolet}, {Genova Santos}, {Gentile}, {George}, {Ghaffari}, {Giacomini},
  {Gianotti}, {Gibb}, {Gillard}, {Gillis}, {Ginolfi}, {Giocoli}, {Girardi},
  {Giri}, {Goh}, {G{\'o}mez-Alvarez}, {Gonzalez}, {Gonzalez}, {Gonzalez},
  {Gouyou Beauchamps}, {Gozaliasl}, {Gracia-Carpio}, {Grandis}, {Granett},
  {Granvik}, {Grazian}, {Gregorio}, {Grenet}, {Grillo}, {Grupp}, {Gruppioni},
  {Gruppuso}, {Guerbuez}, {Guerrini}, {Guidi}, {Guillard}, {Gutierrez},
  {Guttridge}, {Guzzo}, {Gwyn}, {Haapala}, {Haase}, {Haddow}, {Hailey}, {Hall},
  {Hall}, {Hamaus}, {Haridasu}, {Harnois-D{\'e}raps}, {Harper}, {Hartley},
  {Hasinger}, {Hassani}, {Hatch}, {Haugan}, {H{\"a}u{\ss}ler}, {Heavens},
  {Heisenberg}, {Helmi}, {Helou}, {Hemmati}, {Henares}, {Herent},
  {Hern{\'a}ndez-Monteagudo}, {Heuberger}, {Hewett}, {Heydenreich},
  {Hildebrandt}, {Hirschmann}, {Hjorth}, {Hoar}, {Hoekstra}, {Holland},
  {Holliman}, {Holmes}, {Hook}, {Horeau}, {Hormuth}, {Hornstrup}, {Hosseini},
  {Hu}, {Hudelot}, {Hudson}, {Huertas-Company}, {Huff}, {Hughes}, {Humphrey},
  {Hunt}, {Huynh}, {Ibata}, {Ichikawa}, {Iglesias-Groth}, {Ilbert}, {Ili{\'c}},
  {Ingoglia}, {Iodice}, {Israel}, {Israelsson}, {Izzo}, {Jablonka}, {Jackson},
  {Jacobson}, {Jafariyazani}, {Jahnke}, {Jansen}, {Jarvis}, {Jasche}, {Jauzac},
  {Jeffrey}, {Jhabvala}, {Jimenez-Teja}, {Jimenez Mu{\~n}oz}, {Joachimi},
  {Johansson}, {Joudaki}, {Jullo}, {Kajava}, {Kang}, {Kannawadi}, {Kansal},
  {Karagiannis}, {K{\"a}rcher}, {Kashlinsky}, {Kazandjian}, {Keck},
  {Keih{\"a}nen}, {Kerins}, {Kermiche}, {Khalil}, {Kiessling}, {Kiiveri},
  {Kilbinger}, {Kim}, {King}, {Kirkpatrick}, {Kitching}, {Kluge}, {Knabenhans},
  {Knapen}, {Knebe}, {Kneib}, {Kohley}, {Koopmans}, {Koskinen}, {Koulouridis},
  {Kou}, {Kov{\'a}cs}, {Kova\{{\v{c}}\}i{\'c}}, {Kowalczyk}, {Koyama},
  {Kraljic}, {Krause}, {Kruk}, {Kubik}, {Kuchner}, {Kuijken}, {K{\"u}mmel},
  {Kunz}, {Kurki-Suonio}, {Lacasa}, {Lacey}, {La Franca}, {Lagarde}, {Lahav},
  {Laigle}, {La Marca}, {La Marle}, {Lamine}, {Lam}, {Lan{\c{c}}on}, {Landt},
  {Langer}, {Lapi}, {Larcheveque}, {Larsen}, {Lattanzi}, {Laudisio}, {Laugier},
  {Laureijs}, {Lavaux}, {Lawrenson}, {Lazanu}, {Lazeyras}, {Le Boulc'h}, {Le
  Brun}, {Le Brun}, {Leclercq}, {Lee}, {Le Graet}, {Legrand}, {Leirvik}, {Le
  Jeune}, {Lembo}, {Le Mignant}, {Lepinzan}, {Lepori}, {Lesci}, {Lesgourgues},
  {Leuzzi}, {Levi}, {Liaudat}, {Libet}, {Liebing}, {Ligori}, {Lilje}, {Lin},
  {Linde}, {Linder}, {Lindholm}, {Linke}, {Li}, {Liu}, {Lloro}, {Lobo},
  {Lodieu}, {Lombardi}, {Lombriser}, {Lonare}, {Longo}, {L{\'o}pez-Caniego},
  {Lopez Lopez}, {Alvarez}, {Loureiro}, {Loveday}, {Lusso}, {Macias-Perez},
  {Maciaszek}, {Magliocchetti}, {Magnard}, {Magnier}, {Magro}, {Mahler},
  {Mainetti}, {Maino}, {Maiorano}, {Maiorano}, {Malavasi}, {Mamon}, {Mancini},
  {Mandelbaum}, {Manera}, {Manj{\'o}n-Garc{\'\i}a}, {Mannucci}, {Mansutti},
  {Manteiga Outeiro}, {Maoli}, {Maraston}, {Marcin}, {Marcos-Arenal},
  {Margalef-Bentabol}, {Marggraf}, {Marinucci}, {Marinucci}, {Markovic},
  {Marleau}, {Marpaud}, {Martignac}, {Mart{\'\i}n-Fleitas}, {Martin-Moruno},
  {Martin}, {Martinelli}, {Martinet}, {Martin}, {Martins}, {Marulli},
  {Massari}, {Massey}, {Masters}, {Matarrese}, {Matsuoka}, {Matthew},
  {Maughan}, {Mauri}, {Maurin}, {Maurogordato}, {McCarthy}, {McConnachie},
  {McCracken}, {McDonald}, {McEwen}, {McPartland}, {Medinaceli}, {Mehta},
  {Mei}, {Melchior}, {Melin}, {M{\'e}nard}, {Mendes}, {Mendez-Abreu},
  {Meneghetti}, {Mercurio}, {Merlin}, {Metcalf}, {Meylan}, {Migliaccio},
  {Mignoli}, {Miller}, {Miluzio}, {Milvang-Jensen}, {Mimoso}, {Miquel},
  {Miyatake}, {Mobasher}, {Mohr}, {Monaco}, {Mongui{\'o}}, {Montoro}, {Mora},
  {Moradinezhad Dizgah}, {Moresco}, {Moretti}, {Morgante}, {Morisset},
  {Moriya}, {Morris}, {Mortlock}, {Moscardini}, {Mota}, {Moustakas}, {Moutard},
  {M{\"u}ller}, {Munari}, {Murphree}, {Murray}, {Murray}, {Musi}, {Nadathur},
  {Nagam}, {Nagao}, {Naidoo}, {Nakajima}, {Nally}, {Natoli}, {Navarro-Alsina},
  {Navarro Girones}, {Neissner}, {Nersesian}, {Nesseris}, {Nguyen-Kim},
  {Nicastro}, {Nichol}, {Nielbock}, {Niemi}, {Nieto}, {Nilsson}, {Noller},
  {Norberg}, {Nourizonoz}, {Ntelis}, {Nucita}, {Nugent}, {Nunes}, {Nutma},
  {Ocampo}, {Odier}, {Oesch}, {Oguri}, {Magalhaes Oliveira}, {Onoue},
  {Oosterbroek}, {Oppizzi}, {Ordenovic}, {Osato}, {Pacaud}, {Pace}, {Padilla},
  {Paech}, {Pagano}, {Page}, {Palazzi}, {Paltani}, {Pamuk}, {Pandolfi},
  {Paoletti}, {Paolillo}, {Papaderos}, {Pardede}, {Parimbelli}, {Parmar},
  {Partmann}, {Pasian}, {Passalacqua}, {Paterson}, {Patrizii}, {Pattison},
  {Paulino-Afonso}, {Paviot}, {Peacock}, {Pearce}, {Pedersen}, {Peel},
  {Peletier}, {Pellejero Ibanez}, {Pello}, {Penny}, {Percival},
  {Perez-Garrido}, {Perotto}, {Pettorino}, {Pezzotta}, {Pezzuto}, {Philippon},
  {Piersanti}, {Pietroni}, {Piga}, {Pilo}, {Pires}, {Pisani}, {Pizzella},
  {Pizzuti}, {Plana}, {Polenta}, {Pollack}, {Poncet}, {P{\"o}ntinen}, {Pool},
  {Popa}, {Popa}, {Popp}, {Porciani}, {Porth}, {Potter}, {Poulain},
  {Pourtsidou}, {Pozzetti}, {Prandoni}, {Pratt}, {Prezelus}, {Prieto}, {Pugno},
  {Quai}, {Quilley}, {Racca}, {Raccanelli}, {R{\'a}cz}, {Radinovi{\'c}},
  {Radovich}, {Ragagnin}, {Ragnit}, {Raison}, {Ramos-Chernenko}, {Ranc},
  {Raylet}, {Rebolo}, {Refregier}, {Reimberg}, {Reiprich}, {Renk}, {Renzi},
  {Retre}, {Revaz}, {Reyl{\'e}}, {Reynolds}, {Rhodes}, {Ricci}, {Ricci},
  {Riccio}, {Ricken}, {Rissanen}, {Risso}, {Rix}, {Robin}, {Rocca-Volmerange},
  {Rocci}, {Rodenhuis}, {Rodighiero}, {Rodriguez Monroy}, {Rollins},
  {Romanello}, {Roman}, {Romelli}, {Romero-Gomez}, {Roncarelli}, {Rosati},
  {Rosset}, {Rossetti}, {Roster}, {Rottgering}, {Rozas-Fern{\'a}ndez}, {Ruane},
  {Rubino-Martin}, {Rudolph}, {Ruppin}, {Rusholme}, {Sacquegna},
  {S{\'a}ez-Casares}, {Saga}, {Saglia}, {Sahl{\'e}n}, {Saifollahi}, {Sakr},
  {Salvalaggio}, {Salvaterra}, {Salvati}, {Salvato}, {Salvignol},
  {S{\'a}nchez}, {Sanchez}, {Sanders}, {Sapone}, {Saponara}, {Sarpa}, {Sarron},
  {Sartori}, {Sassolas}, {Sauniere}, {Sauvage}, {Sawicki}, {Scaramella},
  {Scarlata}, {Scharr{\'e}}, {Schaye}, {Schewtschenko}, {Schindler},
  {Schinnerer}, {Schirmer}, {Schmidt}, {Schmidt}, {Schmidt}, {Schneider},
  {Schneider}, {Schneider}, {Sch{\"o}neberg}, {Schrabback}, {Schultheis},
  {Schulz}, {Schwartz}, {Sciotti}, {Scodeggio}, {Scognamiglio}, {Scott},
  {Scottez}, {Secroun}, {Sefusatti}, {Seidel}, {Seiffert}, {Sellentin},
  {Selwood}, {Semboloni}, {Sereno}, {Serjeant}, {Serrano}, {Shankar},
  {Sharples}, {Short}, {Shulevski}, {Shuntov}, {Sias}, {Sikkema}, {Silvestri},
  {Simon}, {Sirignano}, {Sirri}, {Skottfelt}, {Slezak}, {Sluse}, {Smith},
  {Smith}, {Smith}, {Smit}, {Soldano}, {Solheim}, {Sorce}, {Sorrenti},
  {Soubrie}, {Spinoglio}, {Spurio Mancini}, {Stadel}, {Stagnaro}, {Stanco},
  {Stanford}, {Starck}, {Stassi}, {Steinwagner}, {Stern}, {Stone}, {Strada},
  {Strafella}, {Stramaccioni}, {Surace}, {Sureau}, {Suyu}, {Swindells},
  {Szafraniec}, {Szapudi}, {Taamoli}, {Talia}, {Tallada-Cresp{\'\i}},
  {Tanidis}, {Tao}, {Tarr{\'\i}o}, {Tavagnacco}, {Taylor}, {Taylor}, {Taylor},
  {Teixeira}, {Tenti}, {Teodoro Idiago}, {Teplitz}, {Tereno}, {Tessore},
  {Testa}, {Testera}, {Tewes}, {Teyssier}, {Theret}, {Thizy}, {Thomas}, {Toba},
  {Toft}, {Toledo-Moreo}, {Tolstoy}, {Tommasi}, {Torbaniuk}, {Torradeflot},
  {Tortora}, {Tosi}, {Tosti}, {Trifoglio}, {Troja}, {Trombetti}, {Tronconi},
  {Tsedrik}, {Tsyganov}, {Tucci}, {Tutusaus}, {Uhlemann}, {Ulivi}, {Urbano},
  {Vacher}, {Vaillon}, {Valdes}, {Valentijn}, {Valenziano}, {Valieri},
  {Valiviita}, {Van den Broeck}, {Vassallo}, {Vavrek}, {Venemans}, {Venhola},
  {Ventura}, {Verdoes Kleijn}, {Vergani}, {Verma}, {Vernizzi}, {Veropalumbo},
  {Verza}, {Vescovi}, {Vibert}, {Viel}, {Vielzeuf}, {Viglione}, {Viitanen},
  {Villaescusa-Navarro}, {Vinciguerra}, {Visticot}, {Voggel}, {von
  Wietersheim-Kramsta}, {Vriend}, {Wachter}, {Walmsley}, {Walth}, {Walton},
  {Walton}, {Wander}, {Wang}, {Wang}, {Weaver}, {Weller}, {Whalen}, {Wiesmann},
  {Wilde}, {Williams}, {Winther}, {Wittje}, {Wong}, {Wright}, {Yankelevich},
  {Yeung}, {Youles}, {Yung}, {Zacchei}, {Zalesky}, {Zamorani}, {Zamorano
  Vitorelli}, {Zanoni Marc}, {Zennaro}, {Zerbi}, {Zinchenko}, {Zoubian},
  {Zucca}, \& {Zumalacarregui}}]{Mellier2024}
{Euclid Collaboration: Mellier}, Y., {Abdurro'uf}, {Acevedo Barroso}, J.~A.,
  {et~al.} 2024, \aap, submitted, arXiv:2405.13491

\bibitem[{{Faber} {et~al.}(1997){Faber}, {Tremaine}, {Ajhar}, {Byun},
  {Dressler}, {Gebhardt}, {Grillmair}, {Kormendy}, {Lauer}, \&
  {Richstone}}]{Faber1997}
{Faber}, S.~M., {Tremaine}, S., {Ajhar}, E.~A., {et~al.} 1997, \aj, 114, 1771

\bibitem[{{Graham} {et~al.}(2003){Graham}, {Erwin}, {Trujillo}, \& {Asensio
  Ramos}}]{Graham2003}
{Graham}, A.~W., {Erwin}, P., {Trujillo}, I., \& {Asensio Ramos}, A. 2003, \aj,
  125, 2951

\bibitem[{{Gualandris} \& {Merritt}(2008)}]{GualandrisMerritt2008}
{Gualandris}, A. \& {Merritt}, D. 2008, \apj, 678, 780

\bibitem[{{Khonji} {et~al.}(2024){Khonji}, {Gualandris}, {Read}, \&
  {Dehnen}}]{Khonji2024}
{Khonji}, N., {Gualandris}, A., {Read}, J.~I., \& {Dehnen}, W. 2024, \apj, 974,
  204

\bibitem[{{Kluge} \& {Bender}(2023)}]{Kluge2023}
{Kluge}, M. \& {Bender}, R. 2023, \apjs, 267, 41

\bibitem[{{Kluge} {et~al.}(2024){Kluge}, {Hatch}, {Montes}, {Golden-Marx},
  {Gonzalez}, {Cuillandre}, {Bolzonella}, {Lan{\c{c}}on}, {Laureijs},
  {Saifollahi}, {Schirmer}, {Stone}, {Boselli}, {Cantiello}, {Sorce},
  {Marleau}, {Duc}, {Sola}, {Urbano}, {Ahad}, {Bah{\'e}}, {Bamford},
  {Bellhouse}, {Buitrago}, {Dimauro}, {Durret}, {Ellien}, {Jimenez-Teja},
  {Slezak}, {Aghanim}, {Altieri}, {Andreon}, {Auricchio}, {Baldi}, {Balestra},
  {Bardelli}, {Bender}, {Bonino}, {Branchini}, {Brescia}, {Brinchmann},
  {Camera}, {Candini}, {Capobianco}, {Carbone}, {Carretero}, {Casas},
  {Castellano}, {Cavuoti}, {Cimatti}, {Congedo}, {Conselice}, {Conversi},
  {Copin}, {Courbin}, {Courtois}, {Cropper}, {Da Silva}, {Degaudenzi}, {Dinis},
  {Duncan}, {Dupac}, {Dusini}, {Farina}, {Farrens}, {Ferriol}, {Fosalba},
  {Frailis}, {Franceschi}, {Fumana}, {Galeotta}, {Garilli}, {Gillard},
  {Gillis}, {Giocoli}, {G{\'o}mez-Alvarez}, {Granett}, {Grazian}, {Grupp},
  {Guzzo}, {Haugan}, {Hoar}, {Hoekstra}, {Holmes}, {Hook}, {Hormuth},
  {Hornstrup}, {Hudelot}, {Jahnke}, {Keih{\"a}nen}, {Kermiche}, {Kiessling},
  {Kitching}, {Kohley}, {Kubik}, {K{\"u}mmel}, {Kunz}, {Kurki-Suonio}, {Lahav},
  {Ligori}, {Lilje}, {Lindholm}, {Lloro}, {Maiorano}, {Mansutti}, {Marggraf},
  {Markovic}, {Martinet}, {Marulli}, {Massey}, {Maurogordato}, {McCracken},
  {Medinaceli}, {Mei}, {Melchior}, {Mellier}, {Meneghetti}, {Merlin}, {Meylan},
  {Moresco}, {Moscardini}, {Munari}, {Nichol}, {Niemi}, {Nightingale},
  {Padilla}, {Paltani}, {Pasian}, {Pedersen}, {Percival}, {Pettorino}, {Pires},
  {Polenta}, {Poncet}, {Popa}, {Pozzetti}, {Racca}, {Raison}, {Rebolo},
  {Renzi}, {Rhodes}, {Riccio}, {Rix}, {Romelli}, {Roncarelli}, {Rossetti},
  {Saglia}, {Sapone}, {Sartoris}, {Sauvage}, {Scaramella}, {Schneider},
  {Schrabback}, {Secroun}, {Seidel}, {Seiffert}, {Serrano}, {Sirignano},
  {Sirri}, {Skottfelt}, {Stanco}, {Tallada-Cresp{\'\i}}, {Taylor}, {Teplitz},
  {Tereno}, {Toledo-Moreo}, {Torradeflot}, {Tutusaus}, {Valentijn},
  {Valenziano}, {Vassallo}, {Verdoes Kleijn}, {Veropalumbo}, {Wang}, {Weller},
  {Williams}, {Zamorani}, {Zucca}, {Biviano}, {Burigana}, {De Lucia}, {George},
  {Scottez}, {Simon}, {Mora}, {Mart{\'\i}n-Fleitas}, {Ruppin}, \&
  {Scott}}]{Kluge2024}
{Kluge}, M., {Hatch}, N.~A., {Montes}, M., {et~al.} 2024, \aap, submitted,
  arXiv:2405.13503

\bibitem[{{Laine} {et~al.}(2003){Laine}, {van der Marel}, {Lauer}, {Postman},
  {O'Dea}, \& {Owen}}]{Laine2003}
{Laine}, S., {van der Marel}, R.~P., {Lauer}, T.~R., {et~al.} 2003, \aj, 125,
  478

\bibitem[{{Lipka} \& {Thomas}(2021)}]{LipkaThomas2021}
{Lipka}, M. \& {Thomas}, J. 2021, \mnras, 504, 4599

\bibitem[{{Lipka} {et~al.}(2024){Lipka}, {Thomas}, {Saglia}, {Bender},
  {Fabricius}, \& {Partmann}}]{Lipka2024}
{Lipka}, M., {Thomas}, J., {Saglia}, R., {et~al.} 2024, arXiv e-prints,
  arXiv:2409.11458

\bibitem[{{Magorrian}(1999)}]{Magorrian1999}
{Magorrian}, J. 1999, \mnras, 302, 530

\bibitem[{{Maraston}(2005)}]{Maraston2005}
{Maraston}, C. 2005, \mnras, 362, 799

\bibitem[{{Marleau} {et~al.}(2024){Marleau}, {Cuillandre}, {Cantiello},
  {Carollo}, {Duc}, {Habas}, {Hunt}, {Jablonka}, {Mirabile}, {Mondelin},
  {Poulain}, {Saifollahi}, {S{\'a}nchez-Janssen}, {Sola}, {Urbano},
  {Z{\"o}ller}, {Bolzonella}, {Lan{\c{c}}on}, {Laureijs}, {Marchal},
  {Schirmer}, {Stone}, {Boselli}, {Ferr{\'e}-Mateu}, {Hatch}, {Kluge},
  {Montes}, {Sorce}, {Tortora}, {Venhola}, {Golden-Marx}, {Aghanim}, {Amara},
  {Andreon}, {Auricchio}, {Baldi}, {Balestra}, {Bardelli}, {Battaglia},
  {Bender}, {Bodendorf}, {Branchini}, {Brescia}, {Brinchmann}, {Camera},
  {Candini}, {Capobianco}, {Carbone}, {Carretero}, {Casas}, {Castellano},
  {Cavuoti}, {Cimatti}, {Congedo}, {Conselice}, {Conversi}, {Copin}, {Courbin},
  {Courtois}, {Cropper}, {Da Silva}, {Degaudenzi}, {Di Giorgio}, {Dinis},
  {Douspis}, {Duncan}, {Dupac}, {Dusini}, {Ealet}, {Farina}, {Farrens},
  {Ferriol}, {Fosalba}, {Fotopoulou}, {Frailis}, {Franceschi}, {Fumana},
  {Galeotta}, {Garilli}, {Gillard}, {Gillis}, {Giocoli}, {G{\'o}mez-Alvarez},
  {Grazian}, {Grupp}, {Guzzo}, {Hailey}, {Haugan}, {Hoar}, {Hoekstra},
  {Holmes}, {Hook}, {Hormuth}, {Hornstrup}, {Hu}, {Hudelot}, {Jahnke},
  {Jhabvala}, {Keih{\"a}nen}, {Kermiche}, {Kiessling}, {Kitching}, {Kohley},
  {Kubik}, {Kuijken}, {K{\"u}mmel}, {Kunz}, {Kurki-Suonio}, {Lahav}, {Le
  Mignant}, {Ligori}, {Lilje}, {Lindholm}, {Lloro}, {Maino}, {Maiorano},
  {Mansutti}, {Marggraf}, {Markovic}, {Martinet}, {Marulli}, {Massey},
  {Maurogordato}, {McCracken}, {Medinaceli}, {Mei}, {Mellier}, {Meneghetti},
  {Merlin}, {Meylan}, {Moresco}, {Moscardini}, {Munari}, {Nakajima}, {Nichol},
  {Niemi}, {Padilla}, {Paltani}, {Pasian}, {Pedersen}, {Percival}, {Pettorino},
  {Pires}, {Polenta}, {Poncet}, {Popa}, {Pozzetti}, {Raison}, {Rebolo},
  {Refregier}, {Renzi}, {Rhodes}, {Riccio}, {Rix}, {Romelli}, {Roncarelli},
  {Rossetti}, {Saglia}, {Sapone}, {Scaramella}, {Schneider}, {Secroun},
  {Seidel}, {Seiffert}, {Serrano}, {Sirignano}, {Sirri}, {Stanco},
  {Tallada-Cresp{\'\i}}, {Taylor}, {Teplitz}, {Tereno}, {Toledo-Moreo},
  {Tsyganov}, {Tutusaus}, {Valentijn}, {Valenziano}, {Vassallo}, {Verdoes
  Kleijn}, {Veropalumbo}, {Wang}, {Weller}, {Williams}, {Zamorani}, {Zucca},
  {Baccigalupi}, {Biviano}, {Burigana}, {De Lucia}, {George}, {Scottez},
  {Viel}, {Simon}, {Mora}, {Mart{\'\i}n-Fleitas}, \& {Scott}}]{Marleau2024}
{Marleau}, F.~R., {Cuillandre}, J.~C., {Cantiello}, M., {et~al.} 2024, \aap,
  submitted, arXiv:2405.13502

\bibitem[{{Martizzi} {et~al.}(2012){Martizzi}, {Teyssier}, \&
  {Moore}}]{Martizzi2012}
{Martizzi}, D., {Teyssier}, R., \& {Moore}, B. 2012, \mnras, 420, 2859

\bibitem[{{McBride} \& {McCourt}(2014)}]{McBride2014}
{McBride}, J. \& {McCourt}, M. 2014, \mnras, 442, 838

\bibitem[{{Mehrgan} {et~al.}(2019){Mehrgan}, {Thomas}, {Saglia}, {Mazzalay},
  {Erwin}, {Bender}, {Kluge}, \& {Fabricius}}]{Mehrgan2019}
{Mehrgan}, K., {Thomas}, J., {Saglia}, R., {et~al.} 2019, \apj, 887, 195

\bibitem[{{Mehrgan} {et~al.}(2023){Mehrgan}, {Thomas}, {Saglia}, {Parikh}, \&
  {Bender}}]{Mehrgan2023}
{Mehrgan}, K., {Thomas}, J., {Saglia}, R., {Parikh}, T., \& {Bender}, R. 2023,
  \apj, 948, 79

\bibitem[{{Mehrgan} {et~al.}(2024){Mehrgan}, {Thomas}, {Saglia}, {Parikh},
  {Neureiter}, {Erwin}, \& {Bender}}]{Mehrgan2024}
{Mehrgan}, K., {Thomas}, J., {Saglia}, R., {et~al.} 2024, \apj, 961, 127

\bibitem[{{Milosavljevi{\'c}} \& {Merritt}(2001)}]{Milosavlijevic2001}
{Milosavljevi{\'c}}, M. \& {Merritt}, D. 2001, \apj, 563, 34

\bibitem[{{Naab} {et~al.}(2009){Naab}, {Johansson}, \& {Ostriker}}]{Naab2009}
{Naab}, T., {Johansson}, P.~H., \& {Ostriker}, J.~P. 2009, \apjl, 699, L178

\bibitem[{{Nasim} {et~al.}(2021){Nasim}, {Gualandris}, {Read}, {Antonini},
  {Dehnen}, \& {Delorme}}]{Nasim2021}
{Nasim}, I.~T., {Gualandris}, A., {Read}, J.~I., {et~al.} 2021, \mnras, 502,
  4794

\bibitem[{{Neureiter} {et~al.}(2023){Neureiter}, {Thomas}, {Rantala}, {Naab},
  {Mehrgan}, {Saglia}, {de Nicola}, \& {Bender}}]{Neureiter2023b}
{Neureiter}, B., {Thomas}, J., {Rantala}, A., {et~al.} 2023, \apj, 950, 15

\bibitem[{{Neureiter} {et~al.}(2021){Neureiter}, {Thomas}, {Saglia}, {Bender},
  {Finozzi}, {Krukau}, {Naab}, {Rantala}, \& {Frigo}}]{Neureiter2021}
{Neureiter}, B., {Thomas}, J., {Saglia}, R., {et~al.} 2021, \mnras, 500, 1437

\bibitem[{{Park} {et~al.}(2017){Park}, {Yang}, {Oonk}, \& {Paragi}}]{Park2017}
{Park}, S., {Yang}, J., {Oonk}, J.~B.~R., \& {Paragi}, Z. 2017, \mnras, 465,
  3943

\bibitem[{{Rantala} {et~al.}(2018){Rantala}, {Johansson}, {Naab}, {Thomas}, \&
  {Frigo}}]{Rantala2018}
{Rantala}, A., {Johansson}, P.~H., {Naab}, T., {Thomas}, J., \& {Frigo}, M.
  2018, \apj, 864, 113

\bibitem[{{Rantala} {et~al.}(2019){Rantala}, {Johansson}, {Naab}, {Thomas}, \&
  {Frigo}}]{Rantala2019}
{Rantala}, A., {Johansson}, P.~H., {Naab}, T., {Thomas}, J., \& {Frigo}, M.
  2019, \apjl, 872, L17

\bibitem[{{Rusli} {et~al.}(2013{\natexlab{a}}){Rusli}, {Erwin}, {Saglia},
  {Thomas}, {Fabricius}, {Bender}, \& {Nowak}}]{Rusli2013}
{Rusli}, S.~P., {Erwin}, P., {Saglia}, R.~P., {et~al.} 2013{\natexlab{a}}, \aj,
  146, 160

\bibitem[{{Rusli} {et~al.}(2013{\natexlab{b}}){Rusli}, {Thomas}, {Saglia},
  {Fabricius}, {Erwin}, {Bender}, {Nowak}, {Lee}, {Riffeser}, \&
  {Sharp}}]{Rusli2013b}
{Rusli}, S.~P., {Thomas}, J., {Saglia}, R.~P., {et~al.} 2013{\natexlab{b}},
  \aj, 146, 45

\bibitem[{{Saglia} {et~al.}(2016){Saglia}, {Opitsch}, {Erwin}, {Thomas},
  {Beifiori}, {Fabricius}, {Mazzalay}, {Nowak}, {Rusli}, \&
  {Bender}}]{Saglia2016}
{Saglia}, R.~P., {Opitsch}, M., {Erwin}, P., {et~al.} 2016, \apj, 818, 47

\bibitem[{{Saglia} {et~al.}(2010){Saglia}, {S{\'a}nchez-Bl{\'a}zquez},
  {Bender}, {Simard}, {Desai}, {Arag{\'o}n-Salamanca}, {Milvang-Jensen},
  {Halliday}, {Jablonka}, {Noll}, {Poggianti}, {Clowe}, {De Lucia},
  {Pell{\'o}}, {Rudnick}, {Valentinuzzi}, {White}, \& {Zaritsky}}]{Saglia2010}
{Saglia}, R.~P., {S{\'a}nchez-Bl{\'a}zquez}, P., {Bender}, R., {et~al.} 2010,
  \aap, 524, A6

\bibitem[{{S{\'a}nchez-Bl{\'a}zquez} {et~al.}(2006){S{\'a}nchez-Bl{\'a}zquez},
  {Peletier}, {Jim{\'e}nez-Vicente}, {Cardiel}, {Cenarro},
  {Falc{\'o}n-Barroso}, {Gorgas}, {Selam}, \& {Vazdekis}}]{Sanchez2006}
{S{\'a}nchez-Bl{\'a}zquez}, P., {Peletier}, R.~F., {Jim{\'e}nez-Vicente}, J.,
  {et~al.} 2006, \mnras, 371, 703

\bibitem[{{Teyssier} {et~al.}(2011){Teyssier}, {Moore}, {Martizzi}, {Dubois},
  \& {Mayer}}]{Teyssier2011}
{Teyssier}, R., {Moore}, B., {Martizzi}, D., {Dubois}, Y., \& {Mayer}, L. 2011,
  \mnras, 414, 195

\bibitem[{{Thomas} {et~al.}(2003){Thomas}, {Maraston}, \&
  {Bender}}]{Thomas2003}
{Thomas}, D., {Maraston}, C., \& {Bender}, R. 2003, \mnras, 339, 897

\bibitem[{{Thomas} \& {Lipka}(2022)}]{ThomasLipka2022}
{Thomas}, J. \& {Lipka}, M. 2022, \mnras, 514, 6203

\bibitem[{{Thomas} {et~al.}(2016){Thomas}, {Ma}, {McConnell}, {Greene},
  {Blakeslee}, \& {Janish}}]{Thomas2016}
{Thomas}, J., {Ma}, C.-P., {McConnell}, N.~J., {et~al.} 2016, \nat, 532, 340

\bibitem[{{Thomas} {et~al.}(2014){Thomas}, {Saglia}, {Bender}, {Erwin}, \&
  {Fabricius}}]{Thomas2014}
{Thomas}, J., {Saglia}, R.~P., {Bender}, R., {Erwin}, P., \& {Fabricius}, M.
  2014, \apj, 782, 39

\bibitem[{{Thomas} {et~al.}(2004){Thomas}, {Saglia}, {Bender}, {Thomas},
  {Gebhardt}, {Magorrian}, \& {Richstone}}]{Thomas2004}
{Thomas}, J., {Saglia}, R.~P., {Bender}, R., {et~al.} 2004, \mnras, 353, 391

\bibitem[{{Veale} {et~al.}(2017){Veale}, {Ma}, {Thomas}, {Greene}, {McConnell},
  {Walsh}, {Ito}, {Blakeslee}, \& {Janish}}]{Veale2017}
{Veale}, M., {Ma}, C.-P., {Thomas}, J., {et~al.} 2017, \mnras, 464, 356

\bibitem[{{Zhao}(1996)}]{Zhao1996}
{Zhao}, H. 1996, \mnras, 278, 488

\end{thebibliography}

%
%
\end{document}